\documentclass[11pt]{article}%{report}
\usepackage{fullpage}
\usepackage[affil-it]{authblk}
\usepackage[margin=3.3cm]{geometry}
\usepackage{amsmath}
\usepackage{dsfont}
\usepackage[english]{babel}
\usepackage[applemac]{inputenc}
\usepackage[T1]{fontenc}
\usepackage[active]{srcltx}
\usepackage{amsfonts}
\usepackage{amssymb}
\usepackage{graphicx}
\usepackage{color}
\usepackage{bm}% bold math
\usepackage{bbm}
\usepackage{authblk}
\usepackage{comment}
\usepackage{enumitem}
\usepackage[titletoc,title]{appendix}
\setlist[enumerate,1]{label={(\Alph*)}}

\newcommand\seq[3]{{#1}_{#2}^{#3}}
\newcommand{\bd}{\begin{document}}
\newcommand{\ed}{\end{document}}
\newcommand{\beq}{\begin{equation}}
\newcommand{\eeq}{\end{equation}}
\newcommand{\bef}{\begin{figure}}
\newcommand{\enf}{\end{figure}}
\newcommand{\bea}{\begin{eqnarray}}
\newcommand{\eea}{\end{eqnarray}}
\newcommand{\baR}{\begin{array}}
\newcommand{\eaR}{\end{array}}
\newcommand{\bc}{\begin{center}}
\newcommand{\ec}{\end{center}}
\newcommand{\ben}{\begin{enumerate}}
\newcommand{\een}{\end{enumerate}}
\newcommand{\bit}{\begin{itemize}}
\newcommand{\eit}{\end{itemize}}
\newcommand{\su}{\section}
\newcommand{\ssu}{\subsection}
\newcommand{\sssu}{\subsubsection}
\newcommand\ssssu[1]{\textbf{#1}}
\newcommand{\nid}{\noindent}
\newcommand{\nnb}{\nonumber}
\newcommand\bloc[2]{{\omega}_{#1}^{#2}}
\newcommand\blocp[2]{{\omega'}_{#1}^{#2}}
\newcommand{\un}{\mathbbm{1}}
\newcommand{\setZ}{\mathbbm{Z}}
\newcommand{\setN}{\mathbbm{N}}
\newcommand{\setR}{\mathbbm{R}}
\newcommand{\pare}[1]{\left(#1 \, \right)}
\newcommand{\scal}[2]{\langle #1 \mid #2 \rangle}
\newcommand{\scalm}[2]{\langle #1 \mid #2 \rangle_{\musp}}
\newcommand{\qdf}[3]{\langle #1 \mid #2 \mid #3 \rangle}
\newcommand{\bra}[1]{\left[#1 \, \right]}
\newcommand{\brac}[2]{\left[\, #1 \, | \, #2 \right]}
\newcommand{\Set}[1]{\left\{\, #1 \, \right\}}
\newcommand{\Setc}[2]{\left\{\, #1 \, ; \, #2 \right\}}
\renewcommand{\P}[1]{\cP\bra{#1}}
\newcommand{\Prob}[1]{\mathds{P}\left[\, #1 \, \right]}
\newcommand{\PT}[2]{\pi^{(T)}_{#2}\left[\, #1 \, \right]}
\newcommand{\pT}{\pi^{(T)}}
\newcommand{\Probex}[1]{P^{(\ast)}\left[\, #1 \, \right]}
\newcommand{\probc}{\mathds{P}}
\newcommand{\Probc}[2]{\mathds{P}\bra{#1 \, \left| \, #2 \right.}}
\newcommand{\probcsp}{\mathds{P}^{(sp)}}
\newcommand{\Probcsp}[2]{\mathds{P}^{(sp)}\bra{#1 \, \left| \, #2 \right.}}
\newcommand{\Probca}[2]{\mathds{P}^{(1)}\bra{#1 \, \left| \, #2 \right.}}
\newcommand{\Pnc}[2]{mathds{P}_n\bra{#1 \, \left| \, #2 \right.}}
\newcommand{\Probcex}[2]{P^{(\ast)}\bra{#1 \, \left| \, #2 \right.}}
\newcommand{\Probcp}[2]{P^{(ap)}\bra{#1 \, \left| \, #2 \right.}}
\newcommand{\abs}[1]{\left|\, #1 \, \right|}
\newcommand{\deq}{\stackrel {\rm def}{=}}
\newcommand{\sqsubsetneq}{\stackrel {\tiny\sqsubset}{\tiny \neq}}
\newcommand\moy[1]{\mu\bra{#1}}
\newcommand\dKL[2]{d_{\tiny{KL}}\bra{#1 \, \mid #2}}
\newcommand\cA{{\cal A}}
\newcommand\cB{{\cal B}}
\newcommand\cC{{\cal C}}
\newcommand\cD{{\cal D}}
\newcommand\cN{{\cal N}}
\newcommand\cX{{\cal X}}
\newcommand\cY{{\cal Y}}
\newcommand\Q{\mathcal{Q}}
\newcommand{\ProbC}[1]{\mathds{P}_\cC\bra{#1}}
\renewcommand\H[1]{{\cH^{(#1)}}}
\newcommand\noy[1]{\nu\bra{#1}}
\newcommand\s[1]{{\cS\bra{#1}}}
\newcommand\Xk[1]{X_k({#1})}
\newcommand\Ik[1]{I_k({#1})}
\newcommand\Skj[1]{S_{kj}({#1})}
\newcommand\X[2]{X_{#1}\pare{#2}}
\newcommand\et[2]{\eta_{#1}\pare{#2}}
\newcommand\tO[2]{\tau_{#1}\pare{#2}}
\newcommand\cF{{\cal F}}
\newcommand\cG{{\cal G}}
\newcommand\cI{{\cal I}}
\newcommand\cH{{\cal H}}
\newcommand\cK{{\cal K}}
\newcommand\cL{{\cal L}}
\newcommand\cM{{\cal M}}
\newcommand\cT{{\cal T}}
\newcommand\cS{{\cal S}}
\newcommand\cP{{\cal P}}
\newcommand\cO{{\cal O}}
\newcommand\cU{{\cal U}}
\newcommand\cV{{\cal V}}
\newcommand\cW{{\cal W}}
\newcommand\f{{\bm{f}}}
\newcommand\baf{{\bar{f}}}
\newcommand\bag{{\bar{g}}}
\newcommand\baF{{\bar{F}}}
\newcommand\baG{{\bar{G}}}
\newcommand\bap{{\bar{\psi}}}
\newcommand\baP{{\bar{\Psi}}}
\newcommand\bF{{\bm{F}}}
\newcommand\bo{{\bm{o}}}
\newcommand\blambda{{\bm{\lambda}}}
\newcommand\bbeta{{\bm{\beta}}}
\newcommand\gk[1]{g_k\pare{#1}}
\newcommand\gLk{g_{L,k}}
\newcommand\akj[1]{\alpha_{kj}(#1)}
\newcommand\sif[1]{\seq{\omega}{-\infty}{#1}}
\newcommand\XkD{\Xk{\bloc{0}{D-1}}}
\newcommand\Xkr{X_k^r}
\newcommand\tLk{\tau_{L,k}}
\newcommand\moyc[2]{\mu\bra{#1 \, | \, #2}}
\newcommand\vkspont{V_k^{(sp)}}
\newcommand\Vkspont[1]{V_k^{(sp)}({#1})}
\newcommand\VkL[1]{V_k^{(L)}({#1})}
\newcommand\vkdet{V_k^{(det)}}
\newcommand\Vkdet[1]{V_k^{(det)}({#1})}
\newcommand\Vksyn[1]{V_k^{(syn)}({#1})}
\newcommand\vkext{V_k^{(S)}}
\newcommand\Vkext[1]{V_k^{(S)}({#1})}
\newcommand\Vknoise[1]{V_k^{(noise)}({#1})}
\newcommand\sk[1]{\sigma_k({#1})}
\newcommand\skr{\sigma_k^r\pare{\bloc{0}{D-1}}}
\newcommand{\ent}[1]{\left[\, #1 \, \right]}
\newcommand\vm[1]{var_m\left[#1 \right]}
\newcommand\eg[1]{\stackrel {\rm #1}{=}}
\newcommand\tk[1]{\tau_k\pare{#1}}
\newcommand\tko{\tau_k(t,\omega)}
\newcommand\cBm[1]{\cB^{-1}\pare{#1}}
\newtheorem{theorem}{Theorem}%[CTh]
\newcommand{\bth}{\begin{theorem}}
\newcommand{\enth}{\end{theorem}}
\newcommand\omD[1]{\seq{\bra{\omega^{(#1)}}}{0}{D-1}}
\newcommand\omz[1]{\omega^{(#1)}(D)}
\newcommand\om[1]{\omega^{(#1)}}
\newcommand\Om[1]{\Omega^{(#1)}}
\newcommand\skjq{s_{kj;q}}
\newcommand\ikq{i_{k;q}}
\newcommand\skjqs{p_{kj;q_s}}
\newcommand\ikqs{i_{k;q_s}}
\newcommand\ikqsu{i_{k;q_{s_u}}}
\newcommand\ikqu{i_{k;q_u}}
\newcommand\xkq{x_{k;q}}
\newcommand\skjqv{s_{kj;q_{v}}}
\newcommand\tjr{t_j^{(r)}(\omega)}
\newcommand\dtt{\Phi(t,\omega)}
\newcommand\dts{\Phi(s,\omega)}
\newcommand\dto{e^{\Phi(t-t_0)}}
\newcommand\dtes{e^{\Phi(t-s)}}
\newcommand{\ket}[1]{\mid #1 \rangle}
\newcommand{\brak}[1]{\langle #1 \mid}
\newcommand{\Sum}[2]{\sum_{\tiny{\baR{ccc} #1 \eaR}}^{\tiny{\baR{ccc} #2 \eaR}}}
\newcommand{\zb}{\zeta_{\bbeta}}
\newcommand{\etab}{\eta_{\bbeta}}
\newcommand{\mex}{\mu^{(\ast)}}
\newcommand{\phex}{\phi^{(\ast)}}
\newcommand{\Phex}{\Phi^{(\ast)}}
\newcommand{\Hex}{\cH_{\bbeta'}^{(\ast)}}
\newcommand\p[1]{\cP\bra{#1}}
\newcommand\w[2]{w_{#1}^{(#2)}}
\newcommand\Gb{\cG_\bbeta}
\newcommand\blp[3]{\omega^{#3,(#1)}_{#2}}
\newcommand\blpb[3]{\overline{\omega^{#3,(#1)}_{#2}}}
\newcommand\moyex[1]{\mu^{(\ast)}\bra{#1}}
\newcommand\E[1]{{\cal E}_{#1}}
\newcommand\Gap{\overline{G}}
\newcommand\gap[1]{\overline{g_{#1}}}
\newcommand{\K}[2]{\cK^{(#1)}(#2)}
\newcommand{\Vr}{V_{reset}}
\newcommand{\bcoul}[2]{{\color{#1}{\textbf{#2}}}}
\newcommand{\bruno}[1]{\bcoul{red}{Bruno says: #1}}
\newcommand{\rod}[1]{\bcoul{blue}{Rodrigo says: #1}}
\newcommand{\tf}[2]{t_{#1}^{(#2)}}
\newcommand{\Tf}[1]{\mathcal{T}^{(#1)}(\omega)}
\newcommand{\Exp}[1]{\mathds{E}\left[\, #1 \, \right]}
\newcommand{\Expc}[2]{\mathds{E}\left[\, #1 \, | \, #2 \right]}
\newcommand{\Expsp}[1]{\mathds{E}^{(sp)}\left[\, #1 \, \right]}
\newcommand{\Expcsp}[2]{\mathds{E}^{(sp)}\left[\, #1 \, \mid \, #2 \right]}
\newcommand{\Cov}[1]{Cov\left[\, #1 \, \right]}
\newcommand{\bD}[2]{\underline{D}_{#1}(#2,\omega)}
\newcommand{\hk}{\hat{k}}
\renewcommand{\i}{\mathrm{i}}

\newcommand{\frc}[1]{\left\{\, #1 \, \right\}}
\newcommand{\Cor}[2]{{\cC}^{(sp)}\left[\, #1, #2 \, \right]}
\newcommand{\Covc}[2]{Cov\left[\, #1 \,\, | \, #2 \right]}
\newcommand{\pr}[3]{p_{#1}\pare{#2, \, #3}}
\newcommand\pTo{{\pi_{\omega}^{(T)}}}
\newcommand\tj[2]{t_j^{(#1)}(#2)}
\newcommand\tjro{\tjr{\omega}}
\newcommand\tjk{t_j^{(k)}(\omega)}
\newcommand\Vk[1]{V_k({#1})}
\newcommand\Vlnoise[1]{V_l^{(noise)}({#1})}
\newcommand\Vnoise[1]{V^{(noise)}({#1})}
\newcommand\VkB[1]{V_k^{(B)}({#1})}
\newcommand\dXk[1]{\delta X_k\pare{#1}}
\newcommand\Repk[2]{\cH^{({#1})}_i\pare{#2}}
\newcommand\zXk[1]{X^{(sp)}_k\pare{#1}}
\newcommand\sko[2]{\sigma_k({#1},{#2},\omega)}
\newcommand\sr{\sigma_R}
\newcommand\sdr{\sigma^2_R}
\newcommand\skd[1]{\sigma^2_k({#1})}
\newcommand\skop[1]{\sigma_k({#1},\omega')}
\newcommand\skopd[1]{\sigma^2_k({#1},\omega')}
\newcommand\ta[2]{\tau_{#1}({#2})}
\newcommand\tauk[1]{\ta{k}{#1}}
\newcommand\tkos{\tau_k(s,\omega)}
\newcommand\tkop{\tau_{k'}(t,\omega)}
\newcommand\tkoptp{\tau_{k'}(t',\omega)}
\newcommand\tlo{\tau_l(t,\omega)}
\newcommand\nko{\tau_k(n,\omega)}
\newcommand\nkrom{\tau_k(r-1,\omega)}
\newcommand\nkrm{\tau_k(r-1,.)}
\newcommand\nkop{\tau_k(n,\omega')}
\newcommand\Vkosyn[2]{V_k^{(syn)}({#1},{#2},\omega)}
\newcommand\Vkoext[2]{V_k^{(S)}({#1},{#2},\omega)}
\newcommand\Vkodet[2]{V_k^{(det)}({#1},{#2},\omega)}
\newcommand\Vkonoise[2]{V_k^{(noise)}({#1},{#2},\omega)}
\newcommand\VkoB[2]{V_k^{(B)}({#1},{#2},\omega)}
\newcommand\Vkto{\Vko{t}{t+1}}
\newcommand\Vksto{\Vko{s}{t}}
\newcommand\tmk{\tau_{M,k}}
\newcommand\gmk{g_{M,k}}
\newcommand\dl[1]{\delta_l\left[#1 \right]}
\newcommand\nega[1]{\stackrel {\rm #1}{\neq}}
\newcommand\moyt[1]{\overline{#1}}
\newcommand\Phisp[1]{\Phi^{(sp)}\pare{#1}}
\newcommand\phisp[1]{\phi^{(sp)}\pare{#1}}
\newcommand\phis{\phi^{(sp)}}
\newcommand\phim{\phi^{(Markov)}}
\newcommand\phispk[1]{\phi^{(sp)}_k\pare{#1}}
\newcommand\Zsp[1]{Z^{(sp)}\bra{#1}}
\newcommand\musp{\mu^{(sp)}}
\newcommand\Csp[2]{\mathcal{C}^{(sp)}\bra{#1, #2}}
\newcommand\Chi[2]{\chi_{#1}\pare{#2}}
\newcommand\CHi[2]{\chi_{#1}^{(#2)}}
\newcommand\moysp[1]{\mu^{(sp)}\bra{#1}}
\newcommand\moyspc[2]{\mu^{(sp)}\bra{#1 \, | \, #2}}
\newcommand\tmoy[1]{\tmu\bra{#1}}
\newcommand\tmoyc[2]{\tmu\bra{#1 \, | \, #2}}
\newcommand\dmu[1]{\delta^{(1)}\moy{#1}}
\newcommand\M[1]{\cM{(#1)}}
\newcommand\pM[2]{\cM^{(#1)#2}}
\newcommand\Mc[3]{\cM_{#2,\,#3}{(#1)}}
\newcommand\pc[2]{p_{#1}{(#2)}}
\newcommand\tphi[1]{\tilde{\phi}^{(#1)}}
\newcommand\tfi{\tilde{\phi}}
\newcommand\tphic[3]{\tilde{\phi}^{(#1)}_{#2,\,#3}}
\newcommand\tfc[2]{\tilde{f}_{#1}{(#2)}}
\newcommand\tmu{\tilde{\mu}}
\newcommand\cst[3]{#1_{#2}^{(#3)}}
\newcommand\spont[2]{#1_{(sp)}\bra{#2}}
\newcommand\spontc[3]{#1_{(sp)}\brac{#2}{#3}}
\newcommand\valp[2]{s_{#1}(#2)}
\newcommand\eqm{\stackrel {\rm \musp}{=}}
\newcommand\emp[3]{\pi_{#1}^{(#2)}\pare{#3}}
\newcommand\Int[2]{\bra{#1, \, #2}}
\newcommand\pder[3]{\left. \frac{\partial #1}{\partial #2}   \right|_{#3}}
\newcommand\cR{\mathcal{R}}
\newcommand\Minv{\mathcal{M}_{inv}}
\newcommand{\tbc}{\bruno{TBC \bigskip}}
\newcommand\bam{\bar{m}}
\newcommand\baM{\bar{M}}
\newcommand\baMc{\bar{M}^{(c)}}
\newcommand\baMcd{\bar{M}^{(c)\dag}}
\newcommand\baMe{\bar{M}^{(e)}}
\newcommand\baMed{\bar{M}^{(e)\dag}}
\newcommand{\rpf}{\cL_\phi}
\newcommand{\RPF}[1]{\cL_\phi\bra{#1}}
\newcommand{\Conv}[3]{\pare{#1 \ast #2}\bra{#3}}

\newtheorem{name}{Printed output}
\newenvironment{remark}[1][Remark:]{\begin{trivlist}
\item[\hskip \labelsep {\bfseries #1}]}{\end{trivlist}}

\begin{document}

\title{Large Deviations Properties of Maximum Entropy Markov Chains from Spike Trains.}
\author[1]{Rodrigo Cofr\'{e} \thanks{Electronic address: \texttt{rodrigo.cofre@uv.cl} ; Corresponding author}}
\author[2]{Cesar Maldonado}
\author[3,4]{Fernando Rosas}
\affil[1]{CIMFAV, Facultad de Ingenier\'{i}a, Universidad de Valpara\'{i}so, Valpara\'{i}so, Chile }
\affil[2]{IPICYT/Divisi\'{o}n de Matem\'{a}ticas Aplicadas, San Luis Potos\'{i}}
\affil[3]{Centre of Complexity Science and Department of Mathematics, Imperial College London, London, UK}
\affil[4]{Department of Electrical and Electronic Engineering, Imperial College London, London, UK}
\maketitle

\begin{abstract}

% Affiliations / Addresses (Add [1] after \address if there is only one affiliation.)
%\address{%
%$^{1}$ \quad CIMFAV, Facultad de Ingenier\'{i}a, Universidad de Valpara\'{i}so, Valpara\'{i}so, Chile; rodrigo.cofre@uv.cl\\
%$^{2}$ \quad IPICYT/Divisi\'{o}n de Matem\'{a}ticas Aplicadas, San Luis Potos\'{i}, Mexico; cesar.maldonado@ipicyt.edu.mx\\
%$^{3}$ \quad Centre of Complexity Science and Department of Mathematics, Imperial College London, London, UK\\
%$^{4}$ \quad Department of Electrical and Electronic Engineering, Imperial College London, London, UK}

% Contact information of the corresponding author
%\corres{Correspondence: rodrigo.cofre@uv.cl}

We consider the maximum entropy Markov chain inference approach to characterize the collective statistics of neuronal spike trains, focusing on the statistical properties of the inferred model. We review large deviations techniques useful in this context to describe properties of accuracy and convergence in terms of sampling size. We use these results to study the statistical fluctuation of correlations, distinguishability and irreversibility of maximum entropy Markov chains. We illustrate these applications using simple examples where the large deviation rate function is explicitly obtained for maximum entropy models of relevance in this field.

\end{abstract}


\section{Introduction}

Spiking neuronal networks are perhaps the most sophisticated information processing devices that are available for scientific inquiry. There exists already an understanding of their basic mechanisms and functionality: they are composed by interconnected neurons which fire action potentials (a.k.a. "spikes") collectively in order to accomplish specific tasks e.g. sensory information processing or motor control \cite{rieke-etal:96}. However, the interdependencies in the spiking activity of populations of neurons can be extremely complex. In effect, these interdependencies can involve neighboring or also distant cells, being established either via structural connections, i.e. physical mediums such as synapses, or by functional connections reflected through spike correlations \cite{friston:11}.

Understanding the way in which neuronal networks process information requires disentangling structural and functional connections while clarifying their interplay, which is a challenging but critical issue~\cite{okatan-wilson-etal:05,ganmor-segev-etal:11a}. For this aim, networks of spiking neurons are usually measured using in-vitro or in-vivo multi-eletrode-arrays, which connect neurons to electronic sensors specially designed for spike detection. Recent progress in acquisition techniques allows the simultaneous measurement of the spiking activity from increasingly large populations of neurons, enabling the collection of large amounts of experimental data \cite{ferrea-etal:12}. Prominent examples of spike train recordings have been obtained from vertebrate retina (salamander, rabbit, degus) \cite{schneidman-berry-etal:06,tkacik-marre-etal:13,vasquez-palacios-etal:12} and cat cortex \cite{marre-boustani-etal:09}. 

However, despite the progress in multi-electrode and neuroimaging recording techniques, modeling the collective spike train statistics is still one of the key open challenges in computational neuroscience. Analysis over recorded data has shown that, although the neuronal activity is highly variable (even when presented repeatedly the same stimulus), the statistics of the response is highly structured \cite{croner:93,shadlen-etal:98}. Therefore, it seems that much of the inner dynamics of neuronal networks is encoded in the statistical structure of the spikes. Unfortunately, traditional methods of estimation, inference, and model selection are not well-suited for this scenario since the number of possible binary patterns that a neuronal network can adopt grows exponentially with the size of the population. As a matter of fact, even long experimental recordings usually contain only a small subset of the possible spiking patterns, which makes the empirical frequencies poor estimators for the underlying probability distribution. For practical purposes, this induces dramatic limitations, as standard inference tools become unreliable as soon as the number of considered neurons grows beyond 10~\cite{schneidman-berry-etal:06}.

Given the binary nature of the spiking data, it is natural to relate neuronal networks and digital communication system via Shannon's information theory. A maybe more subtle way of establishing this link is provided by the physics literature that studies stochastic spins systems. In fact, a succession of research efforts has helped develop a framework to study the spike train statistics based on tools of statistical physics, namely the maximum entropy principle (MEP), which provides an intuitive and tractable procedure to build a statistical model for the whole neuronal network. In 2006 Schneidman \textit{et al} \cite{schneidman-berry-etal:06} and Pillow \textit{et al} \cite{pillow-etal:08}, the MEP was used to characterize the spike train statistics of the vertebrate retina responding to natural stimuli, constraining only range one features namely firing rates and instantaneous pairwise interactions. Since then, the MEP approach has become a standard tool to build probability measures in the field of spike train statistics \cite{pillow-etal:08, schneidman-berry-etal:06,tkacik-etal:13,vasquez-palacios-etal:12}. This approach has triggered fruitful analyses of the neural code, including works about criticality \cite{tkacik:15}, redundancy and error correction \cite{tkacik-marre-etal:13} among other intriguing and promising topics. 

Although relatively successful, this approach for linking neuronal populations and statistical mechanics is based on assumptions that go against fundamental biological knowledge. Firstly, these works assume that the spike patterns are statistically independent of past and future activities of the network. In fact, and not surprisingly, there exists strong evidence supporting the facts that memory effects play a major role in spike train statistics \cite{tang-jackson-etal:08, marre-etal:09, vasquez-palacios-etal:12}. Secondly, most of the literature that applies statistical mechanics to analyze neuronal data use tools that assume that the underlying system is in thermodynamic equilibrium. However, it has been recognized that being out-of-equilibrium is one of the distinctive properties of living systems~\cite{schrodinger:44,deem:07, prigogine:62}. Consequently, any statistical description that is consistent with the out-of-equilibrium condition of living neuronal networks should reflect some degree of time asymmetry (i.e. time irreversibility)~\cite{pshi:10}. 

As a way of addressing the above observations, some recent publications study maximum entropy Markov chains (MEMC) based on a variational principle from the thermodynamic formalism of dynamical systems (see for instance~\cite{cofre-cessac:14, vasquez-palacios-etal:12,cofre-cessac:13b}). This framework is an extension of the classic approach based on the MEP that considers correlation of spikes among neurons simultaneously and with different time delays as constraints, being able in this way to account for various memory effects. 

%\textcolor{red}{Los dos proximos parrafos tiene que estar en linea con las discusiones por emails, la seccion 4 y las conclusiones.} \textcolor{green}{Algun voluntario para esto?.} 
Most of the literature in spike train statistics via the MEP pays little attention to the fact that model estimation is done based on finite data (errors due to statistical fluctuations are likely to occur in this context). As the MEP can be seen as an statistical inference procedure it is natural to inquire about the uncertainty (i.e. fluctuations and convergence properties) related to the inferred MEMC, or, in other words, ask for the robustness of the inference as a function of the sampling size of the underlying data set. Quantifying this error is particularly relevant in the light of recent results that suggest that the parameters inferred by the MEP approach in the context of experimental biological recordings are sharply poised at criticality \cite{mora-bialek:11,tkacik-marre-etal:13}. On the other hand once the MEMC has been inferred it is also important to quantify how well a sample of the MEMC reproduce the average values of features of interest and how likely is that a sample of the MEMC produce a "rare" or unlikely event.

%However, a generalized version of the maximum entropy approach  can be framed within the thermodynamic formalism \cite{bowen:98}, which offers a way to build MEMC consistent with constraints provided by data in a principled way through a variational principle \cite{cofre-cessac:14, vasquez-palacios-etal:12}. Viewing the maximum entropy problem from the point of view of the thermodynamic formalism is particularly useful, as set up the conceptual framework to implement large-deviation methods in order to accurately describe statistical fluctuations of features.

In order to provide some first steps in addressing the above issues, this paper studies the MEMC framework using tools from large deviation theory (LDT) \cite{ellis:85,dembozeitouni:10}. We exploit the fact that the average values of features obtained from samples of the MEMC satisfy a large deviation property, and use LDT techniques to estimate their fluctuations in terms of the sampling size. We also show how to compute the rate functions using the tilted transition matrix technique and the G\"{a}rtner-Ellis theorem. It is to be noted that there is a large body of theoretical work linking the maximum entropy principle and large deviations \cite{ellis:85, georgii:03}. However, these techniques have been scarcely used in spike train analysis (only to study the i.i.d case \cite{balasubramanian:97,mastromatteo:11,macke:13,marsili:13}), most likely because of the lack of a suitable introduction of these concepts within the neuroscientific literature. Consequently, another goal of this paper is to provide an accessible introduction of the MEMC and LDT formalisms to the community of computational neuroscience, avoiding some technicalities while preserving the core ideas and intuitions. This article is part of a more ambitious program that attempts to build a more unified theoretical structure and a complete toolbox helpful to approach spike train statistics using the thermodynamic formalism \cite{cofre-cessac:14,cofre-mal:18}. 

The rest of this paper is organized as follows. Section 2 presents the basic definitions and tools needed to apply large deviations techniques further in the paper. In particular, we present the maximum entropy principle framed in the thermodynamic formalism as a variational principle. In section 3 we introduce some basic aspects of the theory of large deviations. In section 4 we focus on the empirical averages of features. We present some examples of relevance in spike train statistics, where we are able to compute explicitly the rate function for each feature in the maximum entropy potential. In section 5 we present further applications of the theory of large deviations in this field with a list of illustrative examples and finally we present our conclusions in section 6.

%\subsubsection{Notation}

%Let us consider a finite network of $N\geq 2$ neurons. We assume that there is a natural time discretization such that at each time step, each neuron emit at most one spike\footnote{There is a minimal amount of time for neurons in which no two spikes can occur called "refractory period", when binning, usually one goes beyond this time and eventually two spikes occur at the same time bin. In these cases the convention is to consider only one spike.}. We denote the \textit{spike-state} of each neuron $\sigma_{k}^{n}=1$ whenever the $k$-th neuron emits a spike at time $n$, and $\sigma_{k}^{n}=0$ otherwise. The spike-state of the entire network at time $n$ will be denoted $\sigma^n := \bra{\sigma_k^n}_{k=1}^{N}$, which we call it a \textit{spiking pattern}. For $n_{1}\leq n_{2}$, we use the notation $\sigma^{n_1,n_2}$ for a \textwhich is an ordered concatenation of spike patterns $\sigma^{n}$. Now, given $T>0$, a \textit{spike train} %is a spike block $\sigma^{t_0,T}$. In practice $t_0 =0$.

%The Maximum Entropy Principle states that the 'best' model of the data is the distribution $q$
%representing the solution to the following problem

%\begin{equation*}
%\begin{aligned}
%& \underset{q}{\text{maximize}}
%& & H(q) \\
%& \text{subject to}
%& & \mathbb{E}_q[f_m] = \mathbb{E}_{\tilde{p}}[f_m], \; m = 1, \ldots, M.
%\end{aligned}
%\end{equation*}

\section{Preliminaries}

This section introduces the general definitions, notations and conventions that are used throughout the paper, providing in turn the necessary background for the unfamiliar reader.

\subsection{Data binarization and spike trains}\label{sec:2.1}

Let us consider a set of measurements from a network of $N$ interacting neurons. The "raw data" consists of $N$ continuous signals containing the extra-cellular potential (electrical potential measured outside of the cell) of each of the neurons recorded over the length of the experiment. This data is processed by spike sorting algorithms \cite{quian:04,hill-etal:11}, which are signal processing techniques designed to extract the spiking activity of each neuron. 

\begin{figure}[h!]
  \centering
\    \includegraphics[width=0.6\textwidth]{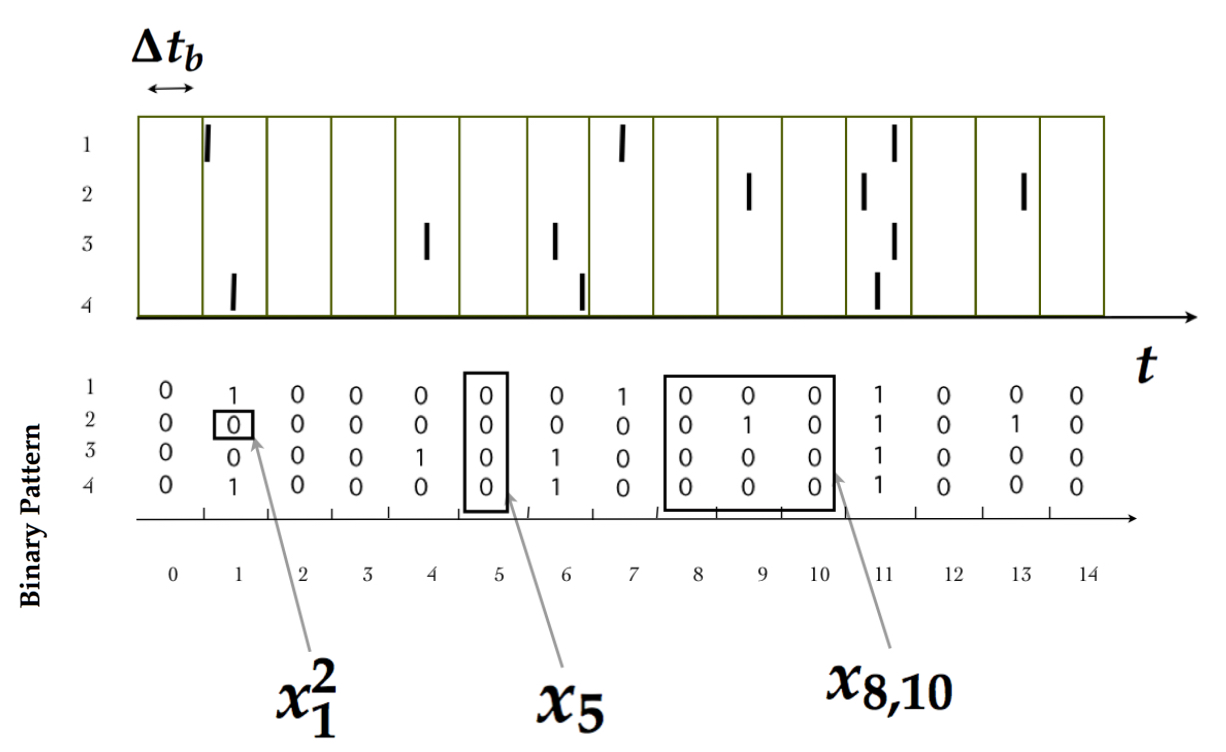}
    \caption{(Top) Each bar indicates a spike of a neuron indexed from 1 to 4 in continuous time. (Bottom) After binning $\Delta t_b$ the spiking activity is transformed into binary patterns in discrete time. We illustrate the notation used throughout this paper. }
     \label{fig:fa}
\end{figure}

Neurons have a minimal characteristic time in which no two spikes can occur, called "refractory period" \cite{gerstner-kistler:02}, which provides a natural time-scale that can be used for "binning" (i.e. for discretizing) the time index of the measurements,  denoted by $\Delta t_b$\footnote{When binning, sometimes can be useful to go beyond the refractory period. In those cases, two spikes may occur within the same time bin. The convention is to consider this event equivalent to just one spike.}. Denoting the time index by the integer variable $t$, one can say that $x^k_t=1$ whenever the $k$-th neuron emits a spike during the $t$-th time bin, while $x^k_t=0$ otherwise. This standard procedure transforms experimental data into sequences of binary patterns (see figure \ref{fig:fa}).

A \textit{spike pattern} is the spike-state of all the measured neurons at time bin $t$, which is denoted by $\boldsymbol{x}_t := \big[x^{k}_t\big]_{k=1}^{N}$. A \textit{spike block} is a consecutive sequence of spike patterns, denoted by $\boldsymbol{x}_{t,r} := \big[\boldsymbol{x}_s\big]_{s=t}^{r}$. While the length of the spike block $\boldsymbol{x}_{t,r}$ is $r-t+1$, is also useful to consider spike blocks of infinite length starting from time $t=0$, which are denoted by $\boldsymbol{x}$. Finally, is this paper we consider that a \textit{spike train} is either a spike block of finite length or an infinite sequence of spiking patterns, which will be useful later when discussing asymptotic properties. The set of all possible spike blocks of length $R$ corresponding to a network of $N$ neurons is denoted by $\mathcal{A}^{N}_R$. The set of all spike blocks of infinite length is denoted by $\Omega \equiv \mathcal{A}_{\mathbb{N}}^N$, which is a useful mathematical object as clarified below. Let us define $proj_R:\Omega \rightarrow \mathcal{A}_R^N$ the natural projection given by $proj_R(\boldsymbol{x}) = \boldsymbol{x}_{0,R-1}$.

\subsection{Features}\label{sec:features}

Following the machine-learning nomenclature, a \textit{feature} is a function that extracts a property of interest from the data. Formally, a \textit{feature} is defined as a function $f:\Omega \rightarrow \mathbb{R}$ that associates a real number to each $\boldsymbol{x}\in\Omega$.  The feature $f$ is said to have a temporal range or simply a range $R$ if for every $\boldsymbol{x},\boldsymbol{y} \in \Omega$ such that $\boldsymbol{x} \neq \boldsymbol{y}$, one has that $f(\boldsymbol{x}) = f(\boldsymbol{y})$ if and only if $\boldsymbol{x}_{0,R-1} = \boldsymbol{y}_{0,R-1}$, that is, if $f$ only depends on the first $R$ spike patterns of the spike-train. A special class of features, over which this work is focused on, are binary functions consisting of finite products of spike states, i.e. 
$$
f_l(\boldsymbol{x})=\prod_{k=1}^q x^{i_k}_{t_k}.
$$
\noindent
Above, $l$ is a shorthand notation for the set $\{t_k,i_k\}_{k=1}^q$, where $[t_k]_{k=1}^q$ and $[i_k]_{k=1}^q$ are collections of time and neuron indexes respectively, and $q$ is the number of spikes considered by the feature. Correspondingly, for a given index $l$, one has $f_l(\boldsymbol{x})=1$ if and only if the $i_k$-th neuron spikes at time $t_k,$ for all $k \in \{1,\dots,q\}$ in the spike-train $\boldsymbol{x}$, while $f_l(\boldsymbol{x})=0$ otherwise. Note that, when considering features of range $R \geq 1$, the firing times $t_k$ are constrained within the interval $\Set{0, \dots, R-1}$. We define the reduced feature $\tilde{f}:\mathcal{A}_\mathbb{N}^R\rightarrow\mathbb{R}$ such that 
$$
\tilde{f}(\boldsymbol{x}_{0,R-1}) = \tilde{f}(proj_R(\boldsymbol{x})) = f(\boldsymbol{x}).
\label{eq:reduced}
$$

\subsection{Statistical structure}\label{StatStruc}

For a given spiking neuronal network involved in a particular experimental protocol, the measured activity usually contains a significant amount of stochasticity that is characteristic of measurements at this spatiotemporal scale. This randomness is caused mostly by
\begin{itemize}
\item[(i)] the random variation in the ionic flux of charges crossing the cellular membrane per unit time at the post synaptic button due to the binding of neurotransmitter, 
\item[(ii)] the fluctuations in the current resulting from the large number of opening and closing of ion channels \cite{schwalger-etal:10,linaro-etal:11},
\item[(iii)] noise coming from electrical synapses \cite{cofre-cessac:13}.
\end{itemize}

In order to capture this stochasticity within our modeling, it is natural to endow $\Omega$ with a probabilistic structure. For this, we assume that exists a probability distribution $p\{\cdot\}$ over $\Omega$ that quantifies the intrinsic randomness that is associated to the spiking phenomena. From this point of view, all $A \in \Omega$ are \textit{events} that might take place with probability $p\{A\}$. Following a standard practice in computational neuroscience, we assume that the stochastic process generating the spikes is stationary i.e. that their statistics do not change in time. As we will discuss below this assumption is crucial for the maximum entropy inference. Although an extension of our approach to a non-stationary scenario is possible, we focus here on the stationary case as it greatly simplifies the presentation. Using the stationary assumption, given the probability distribution of the whole process $p\{\cdot\}$ one can define a unique corresponding probability distribution over $\mathcal{A}^{N}_R$ following the natural projection, given by:
\begin{equation}\label{eq:prob-projection}
p_R\{B\in \mathcal{A}^{N}_R\} := p\{proj_R^{-1}(B) \in \Omega\} .
\end{equation}

\noindent
As a consequence of assuming an stochastic process guiding the neuronal activity, a feature $f:\Omega\rightarrow\mathbb{R}$ becomes a random variable. Consequently, the statistics of $f$ are defined by
$$
p\{ f = a \} = p\{\boldsymbol{x}\in\Omega|f(\boldsymbol{x})=a\}.
$$ 
In particular, considering the feature $f(\boldsymbol{x})=x_t^k$, one can note that individual spike-states (as well as spike patterns and spike blocks) become discrete random variables. As a convention, we denote $X_t^k$ a random spike-state that follows an implicit underlying probability distribution $p\{\cdot\}$, while lower-case expressions (e.g. $x_t^k$) are used for denoting concrete realization of these random variables. The mean value of a feature $f$ with respect to the probability $p\{\cdot\}$ is given by:
$$
\mathbb{E}_p\{f\} = \sum_{\boldsymbol{x}\in\Omega}f(\boldsymbol{x})p\{\boldsymbol{x}\}.
$$
For the case of features of range $R$, the mean value can be expressed alternatively as:
$$
\mathbb{E}_p\{f\} = \sum_{\boldsymbol{x}_{0,R-1}\in \mathcal{A}^{N}_R}\tilde{f}(\boldsymbol{x}_{0,R-1})p_R\{\boldsymbol{x}_{0,R-1}\} = \mathbb{E}_{p_R}\left\{\tilde{f}\right\}
$$
which is a finite sum. Above, $\tilde{f}$ is the reduced feature, as defined in \eqref{eq:reduced}.

\subsection{Empirical averages}

Let us consider spiking data of the form $\boldsymbol{x}_{0,T-1}$, where $T$ is the sample length. Although in general the underlying probability measure $p\{\cdot\}$ that govern the spiking activity is unknown, it is useful to use the available data to estimate the mean values of specific features.  If $f$ is a feature of range $R$, the empirical average value of $f$ from the sample $\boldsymbol{x}_{0,T-1}$ is 
\begin{equation} \label{empirical_average}
A_{T}(f)=\frac{1}{T-R+1}\sum_{i=0}^{T-R} f(\boldsymbol{x}_{i,R-1+i}).
\end{equation}
\noindent
In particular, for features of range one, the previous expression becomes $A_{T}(f) = \frac{1}{T} \sum_{i=0}^{T-1}f(\boldsymbol{x}_i)$. 

An interesting questions is under which conditions $A_T(f)\rightarrow \mathbb{E}_p\{f\}$ as $T$ grows. This, and other convergence issues, are explored in Section~\ref{sec:LDandapp}.

\section{Inference of the statistical model with the MEP}~\label{sec:3}

Following Section~\ref{StatStruc}, the probability measure $p\{\cdot\}$ represents the inherent stochasticity of the spiking neural population under consideration. As $p\{\cdot\}$ is unknown, one would like to infer it from data. In the sequel, we first introduce the general MEP as a method for inferring $p\{\cdot\}$. Then, we show this principle for the case where only synchronous constraints are considered, and finally, we present the case of where non-synchronous correlations are included to constrain the maximization problem.

\subsection{Fundamentals of the MEP}\label{sec:3.1}

The MEP was first proposed by E. T. Jaynes as a way for estimating probability distributions when the information for performing the inference is scarce~\cite{jaynes:57}. Rooted in principles of statistical physics, this approach selects a probability measure that satisfies the evidence supported by the available information while leaving all other aspects as random as possible. For quantifying the corresponding randomness, Jaynes shows that the most natural metric is the Shannon entropy ~\cite{jaynes:03}. The probability measure found by this procedure is known as the \emph{maximum entropy distribution}.  
%consists in performing an optimal inference under incomplete information, as an extension of traditional logic . 

Formally, the MEP is a concave constrained maximization problem, where the constraints that define the optimization space correspond to the available information that guide the inference process. 
%The specific probability distribution that achieves the maximum entropy is then selected. 
Accordingly, if additional constraints are introduced then the optimization space is reduced; this corresponds to the informative power of new information, which restricts the space of models that are consistent with it.

The inference procedure based on the MEP follows the following steps:
\begin{itemize}

\item[I.] Choose $K$ features of interest $f_1,\dots,f_K$ (c.f. Section~\ref{sec:features}). 

\item[II.] Using the available data $\boldsymbol{x}_{0,T-1}$, compute the empirical averange of each feature $A_T(f_k):=c_k$.

\item[III.] Assuming stationarity, define the space of statistical models $\mathcal{M}(c_1,\dots,c_K) \subset \mathcal{M}$ given by
$$
\mathcal{M}(c_1,\dots,c_K) = \{ g \in\mathcal{M} | \; \mathbb{E}_g\{f_1\} = c_1,\dots, \mathbb{E}_g\{f_K\} = c_K \},
$$
where $\mathcal{M}$ is the set of probability measures and $\mathcal{M}(c_1,\dots,c_K)$ is the family of probability measures that are consistent with the empirical mean values $c_1,\dots,c_K$ obtained in Step II.

\item[IV.] Defining the entropy rate of the stochastic process as
\begin{equation}\label{SHE}
\mathcal{S}\{p\}= \lim_{t\rightarrow\infty} \frac{1}{t} \sum_{\boldsymbol{x}_{0,t-1}\in \mathcal{A}_t^N}  p_t\{\boldsymbol{x}_{0,t-1}\} \log   \frac{1}{p_t\{\boldsymbol{x}_{0,t-1}\}},
\end{equation}
\noindent
find the maximum entropy process, characterized by the probability measure
\begin{equation}\label{maxentp}
\hat{p} = \max_{q \,\in\, \mathcal{M}(c_1,\dots,c_k)} %\Big\{
\mathcal{S}\{q\}. %: \mathbb{E}_{\nu}(f_k) =C_k, \quad  \ \forall\ k \in \{ 1,\dots,K\} \Big\}.
\end{equation}

\end{itemize}

Some small remarks are to be said about this procedure. One can think of this as a data-driven algorithm, whose input is the data $\boldsymbol{x}_{0,T-1}$ and output is the maximum entropy measure $\hat{p}$. The first two steps of the process are known in the machine learning literature as "feature selection" and "feature extraction", respectively (see e.g. \cite{peng:05,brown:12}). The goal of these steps is to reduce the dimensionality of the input for the subsequent stages, in order to prevent the selected model to overfitting the data (i.e. to include in the model effects of noise and biases due to the finiteness of the data). Hence, what drives the model selection stages is not the whole data but the quantities $c_1,\dots,c_K$, which define the space to be explored in Step 4. 

Steps 3 and 4 are known as "model selection". According to the the machine learning jargon these steps deliver a generative model, in the sense the obtained model can later be used to generate new data. In this sense, it is interesting that although the data is finite, the entropy rate calculated in Step 4 is computed over all spike blocks of all lengths $t$, which is possible due to the generative nature of the candidate models. The inputs for the model selection stages are not the whole data $\boldsymbol{x}_{0,T-1}$ but only the values $c_1,\dots,c_K$, which represent the knowledge obtained from the data that guides the search in the space of candidate models. Moreover, as these quantities represent all the available knowledge one has about the underlying stochastic process generating the spikes, therefore, one would like to select a model that reflect that information while making no further assumptions. By recalling the work made by Claude Shannon on the analysis of information sources (c.f. ~\cite{cover:06} and references therein), one can interpret the entropy rate as a measure of how hard is to predict the realization of a stochastic process. This implies, in turn, that the maximum entropy measure within $\mathcal{M}\{c_1,\dots,c_K\}$ is the most random, i.e unstructured, among those which satisfies the constraints $A_T(f_1)=c_1,\dots,A_T(f_K)=c_k$. Although the framework presented above is general enough to encompass the cases when considering only synchronous constraints and when considering also non-synchronous constraints, the methods used to find the maximum entropy measure are different. In section \ref{tic} we present the method for finding the maximum entropy measure when only synchronous features are selected, leaving for section \ref{tic} the more general situation including non-synchronous constraints.

%It is important to note that if the features $f_1,\dots,f_K$ are not chosen appropriately, the resulting model space could allow the Shannon entropy in \eqref{maxentp} to be unbounded, and hence in this case $\hat{p}$ could be ill-defined. Appropriate feature selection hence play a key role in delivering a satisfactory model. \textcolor{red}{(are we discussing this further or not? if not, why not? is left for future work?)} %\textcolor{red}{(En muchos schemes de machine learning el demonio esta en pequenos pasos intermedios como este. Seria bueno aclarar la importancia que tiene la feature selection en la obtencion del modelo...)}

\subsection{Time-independent constraints}\label{tic}

Assuming only synchronous interactions is equivalent to only consider features of range one (i.e. features that consider neurons at the same time index, c.f. Section~\ref{sec:features}), which leads to restricting the candidate models to those in where the present state is statistically independent of past and future states. Moreover, by the assumption of stationarity, this leads to modeling the collective spiking activity as a sequence of i.i.d. random variables. The challenge, in this case, is to estimate the corresponding distribution as reliably as possible. 
A large portion of the literature of maximum entropy spike train statistics focus on synchronous interactions between neurons,  implicitly neglecting interactions across time. Although this assumption induces a strong simplification, the resulting models have proved to be rich in structure and can provide interesting results and insights about the neural code ~\cite{pillow-etal:08, schneidman-berry-etal:06}. In the following, we recall how this problem can be addressed using the MEP.

%\subsubsection{Choosing the best fit}

%and with measurements of the average value of $K$ arbitrarily chosen features $\{f_k\}_{k=1}^K$ of range $1$, i.e., time-independent. That is, for each $k=1,\ldots,K$, $A_T(f_k)=C_k$. The $K$ averages $A_T(f_k)$ are the constraints of the maximization problem.
%\noindent

%\noindent
%where $\mathcal{M}$ is the set of probability measures. 

As a consequence of the assumptions of temporal independence and stationarity, it can be shown that \eqref{maxentp} is reduced to
\begin{equation}\label{eq:mep_1}
\hat{p}_1 = \max_{q_1 \,\in\, \mathcal{M}_1(c_1,\dots,c_k)} \sum_{\boldsymbol{x}_0\in\mathcal{A}_1^N} q_1\{\boldsymbol{x}_0\} \log \frac{1}{q_1\{\boldsymbol{x}_0\}}
\end{equation}
where $\mathcal{M}_1(c_1,\dots,c_k)$ corresponds to the set of distributions $q_1$ over $\mathcal{A}_1^N$ (c.f. range one projections in \eqref{eq:prob-projection}) such that the constraints $\mathbb{E}_{q_1}\{f_k\} = c_k$ are satisfied for each $k=1,\dots,K$. Note that the above sum is over the $2^N$ possible spike patterns, being a simpler condition than $\eqref{maxentp}$. In fact, following a simple argument based on Lagrange multipliers and the concavity of the  entropy, it can be show that the distribution $\hat{p}_1$ that solves \eqref{eq:mep_1} is unique. Moreover, is a Boltzmann-Gibbs distribution~\cite{jaynes:03}:
\beq\label{maxentpd}
\hat{p}_1\{\boldsymbol{x}_0\}=\frac{e^{-\mathcal{H}_{\boldsymbol{\beta}}(\boldsymbol{x}_0)}}{Z( \boldsymbol{\beta})} \quad \forall \boldsymbol{x}_0 \in \mathcal{A}_1^N; \quad Z( \boldsymbol{\beta})= \sum_{\boldsymbol{x}_0\in\mathcal{A}_1^N} e^{-\mathcal{H}_{\boldsymbol{\beta}}(\boldsymbol{x}_0)},
\eeq
%\end{itemize}

%The SCGF associated to the i.i.d process is in this case $\log Z(\beta)$. For the observables one can use the contraction principle KL divergence. Otherwise matrix. 
\noindent
where  $\mathcal{H}_{\boldsymbol{\beta}}$ is referred as the \textit{energy or potential} function 

\beq\label{energy}
\mathcal{H}_{\boldsymbol{\beta}}(\boldsymbol{x}_0)=\sum_{k=1}^K \beta_k \tilde{f}_k(\boldsymbol{x}_0),
\eeq
\noindent
$\boldsymbol{\beta} \in \mathbb{R}^K$ is the vector of Lagrange multipliers. Following the statistical physics literature $Z(\boldsymbol{\beta})$ is called the \textit{partition function}, whose logarithm is referred as \textit{free energy}.

Conversely, from the uniqueness property of the maximum entropy distribution one can conclude that there is only one Boltzman-Gibbs distribution $\hat{p}_1$ that belongs to $\mathcal{M}(c_1,\dots,c_K)$, being the only solution of \eqref{eq:mep_1}. Interestingly, this alternative approach is much easier to solve the original optimization problem\footnote{In particular, $\mathcal{M}_1\{c_1,\dots,c_k\}$ is not easy to parametrize and hence the application of standard techniques of convex optimization (e.g. gradient decent) is not straightforward.}. In effect, one only need to find the values of the parameter vector $\beta_k$ that reproduces the empirical average values $c_1,\dots,c_K$. Moreover, it is known that for any Boltzmann-Gibbs distribution $p_1$ the following holds: 
\beq \label{fitr1}
\frac{\partial \ln Z(\boldsymbol{\beta}) }{\partial\beta_k}= \mathbb{E}_{\hat{p}_1}(\tilde{f}_k).
\eeq
Therefore, using \eqref{fitr1} one could find the appropriate values of $\boldsymbol{\beta}$ for which  $\mathbb{E}_{\hat{p}_1}\{\tilde{f}_k\} = c_k$ are satisfied\footnote{However, for practical purposes this problem cannot be solved for systems with $N>20$ \cite{vasquez-palacios-etal:12}, %\textcolor{red}{(REFERENCE) Es tambien raro por que es tan dificil... alguien puede explicar brevemente el por que no se puede? es un probleman numerico? computacional? teorico?} \textcolor{green}{Hay que calcular el valor propio mas grande de una matrix gigantezca, y despues derivarlo con respecto a cada uno de los parametros. Esto se puede hacer explicitamente como como en los ejemplos que mostramos, pero no para matrices muy grandes. La alternativa numerica es posible, es decir, tirar los valores de los parametros al azar e ir variandolos hasta converger. Esto se hace usualmente usando monte carlo, aprovechando metodos de sampleo de tipo Gibbs, nada evidente.}
so alternative procedures are needed. For the interested reader, we refer to \cite{nasser-marre-etal:14, tkacik-prentice-etal:10, tkacik-marre-etal:13, tkacik:15}.}.

\subsection{Non-synchronous constraints}\label{tdc}

%If one assumes that spike trains can be characterized by Markov chains, the empirical sampling problem, mentioned in the previous section becomes even more pronounced. Therefore, the maximum entropy approach becomes an appealing alternative.

%In the following, we present a framework that allows generalizing the MEP to build MEMC that are consistent with available information.
A generalization of the previous approach is to include average values of features corresponding to interactions in the spiking activity across time as constraints. This is a natural assumption in biological spiking networks as it is expected that the spike of one neuron influence other subsequent spikes.
\noindent
Statistical models with time-dependencies can be generated with the MEP by introducing features that involve different time indexes. In effect, selecting features of range $R$ induces interdependencies and a corresponding "memory" in the model of length $R-1$, and thus it is natural to look for the best suited Markov chain over the state space $\mathcal{A}_N^R$.
%Accordingly, one can show that in this case the resulting stochastic process of the MEP procedure described in Section XX is a time-homogeneous Markov chain of order $R-1$, \cite{vasquez-palacios-etal:12,cofre-cessac:14} \footnote{In fact, any additional memory of length beyond $R-1$ would induce interdependencies that are not determined by the constraints, violating the principle of faithfulness to the available knowledge of the MEP.}. 
A Markov chain model would allow to express the probability of a spike train $\boldsymbol{x}_{0,T}$ for $T>R$ as
$$
p\{\boldsymbol{x}_{0,T}\} = 
\pi\{\boldsymbol{x}_{0,R-1}\}
P\{\boldsymbol{x}_{1,R}|\boldsymbol{x}_{0,R-1}\}
\cdots 
P\{\boldsymbol{x}_{T-R,T-1}|\boldsymbol{x}_{T-R+1,T}\},
$$
%where $p$ is a stationary Markov measure one is looking for, which is characterized by
where $P$ is a transition probability matrix \footnote{Note that $P\{\cdot,\cdot\}$ has a consistency requirement: for $\boldsymbol{w},\boldsymbol{y} \in \mathcal{A}_N^R, P\left\lbrace \boldsymbol{w}|\boldsymbol{y} \right\rbrace>0$ only when $\boldsymbol{y}_{1,R-1}=\boldsymbol{w}_{0,R-2}$.} and $\pi$ is a corresponding invariant probability distribution (which is unique due to the ergodicity assumption, c.f. Section \ref{tmm}) associated to $P$. Note that, due to the stationarity condition, the transition probabilities $P\{\cdot|\cdot\}$ are constant over the realization of the whole process (see Appendix~\ref{Appen-Markov-Spike} for more details.). 

Following the MEP as described in Section~\ref{sec:3.1}, we look for a procedure for finding a Markov transition matrix $P$ that maximizes its entropy rate while satisfying some constrains given the empirical averages of observables $f_1,\dots,f_K$. For ergodic Markov chains, a well-known calculation (that can be found e.g. in \cite{cover:06}) shows that the entropy rate, as given by \eqref{SHE}, is equivalent to the following simple expression:
\beq \label{KSE}
\mathcal{S}_{KS}(\pi,P) = -\sum_{ i,j \in \mathcal{A}_N^R} \pi_i \sum_{j} P_{ij} \log  P_{ij} .
\eeq
where $\pi_i=\pi\{\boldsymbol{x}_{0,R-1}=i\}$ and $P_{ij}=P\{j|i\}$ for $i,j\in\mathcal{A}_R^N$. Is important to notice that \eqref{KSE} corresponds to the \textit{Kolmogorov-Sinai entropy (KSE)} in the dynamical systems literature~\cite{sinai:72}. In general \eqref{KSE} is larger when for a fixed $i$ the conditional probabilities $P_{ij}$ are closer to an uniform distribution, i.e. when knowing the transition statistics gives little certainty about the next step.

It can be shown that, if the considered features do not involve correlations across time (i.e. they are features of range 1, c.f. Section~\ref{sec:features}), then the resulting transition probabilities are such that the corresponding stochastic process is i.i.d (i.e. when $P_{ij}=\pi_j$). Interestingly, in this scenario equation \eqref{KSE} reduces to the Shannon entropy of $\pi_i$. This clarifies that this approach based on Markov chains extends the classical MEP and the results presented in Section~\ref{tic}.  

%It is to be noted that
%, although quite similar in spirit to the procedure presented in Section~\ref{sec:3.1}, 
Finding the MEMC raises, however, some extra technicalities with respect to the time-independent case. Recall that the goal is no longer to estimate a probability distribution, but to reconstruct from data a transition matrix $P$ and a corresponding invariant measure $\pi$. %, based on the criterion that the entropy rate of the stochastic process should be maximized. 
The challenge is that as $P$ and $\pi$ are not independent parameters of the process ($\pi$ has to be the eigenvector associated with the unitary eigenvalue of $P$ \cite{chliamovitch:15}), therefore one cannot apply Lagrange multipliers over the entropy rate function \eqref{KSE}. In the sequel we explore an alternative route to build the MEMC based on the transfer matrix technique. This technique is computationally simple, and also provides further insightful connections with statistical physics and thermodynamics.

\subsubsection{Transfer Matrix Method}\label{tmm}

In order to find the MEMC associated with non-synchronous constraints, we follow the same ideas presented above in the time-independent case, but using different tools. We present them here.

Let us consider the set of features chosen to constrain the maximization of entropy rate (step I in \ref{sec:3.1}). We assume that the features chosen have a finite maximum range $R$. From these features one can build the energy function $\mathcal{H}_{\boldsymbol{\beta}}$ (\ref{energy}) of finite range $R$ as a linear combination of these features. Using this energy function we build a matrix denoted by $\mathcal{L}_{\mathcal{H}_{\boldsymbol{\beta}}}$, so that for every $y, w\in \mathcal{A}^{N}_R$ its entries are given as follows:

%The Ruelle-Perron-Frobenius matrix for a finite range energy function $\mathcal{H}_{\boldsymbol{\beta}}$ \textcolor{red}{(que es esto? donde esta definido?)} is denoted by $\mathcal{L}_{\mathcal{H}_{\boldsymbol{\beta}}}$ and defined as follows:

\beq\label{transfmatrix}
\mathcal{L}_{\mathcal{H}_{\boldsymbol{\beta}}}(y,w)=   
\left\{
\begin{array}{lll}
 e^{\mathcal{H}_{\boldsymbol{\beta}}( yw_{R-1})}
\quad &\mbox{if }   y_{1,R-1}= w_{0,R-2}   \\
0, \quad &\mbox{otherwise}.
\end{array}
\right. 
\eeq
%\textcolor{red}{El siguiente texto es poco amable para la audiencia no matematica.} \textcolor{green}{Estoy de acuerdo, sin embargo, creo que es importante regirse por los standares del formalismo termodinamico, es lo que hemos venido haciendo y creo que es la direccion adecuada, pues es la que abre mas nuevas posibilidades de exploracion. Yo aca no me pondria novedoso ni cambiaria la forma de introducir los conceptos, pues son resultados conocidos con notacion standard}
By $yw_{R-1}$ we mean the word obtained by concatenation of $y_1$ and $w_{1,R-1}$. In the particular case of energy functions associated to range one features, we the aboves matrix is defined as $\mathcal{L}_{\mathcal{H}_{\boldsymbol{\beta}}}(y,w)=e^{\mathcal{H}_{\boldsymbol{\beta}}(y)}$. Assuming $\mathcal{H}_{\boldsymbol{\beta}} > -\infty$, the elements of the matrix $\cL_{\mathcal{H}_{\boldsymbol{\beta}}}$ are non-negative, this in turn implies ergodicity. Moreover, the matrix is primitive by construction, thus it satisfies the Perron-Frobenius theorem \cite{seneta:06}. Hereafter $\mathcal{L}_{\mathcal{H}}$ will be referred as the Ruelle-Perron-Frobenius matrix (RPF). Let us denote be $\rho$ the largest eigenvalue of $\mathcal{L}_{\mathcal{H}}$, which because it satisfies the Perron-Frobenius theorem is an eigenvalue of multiplicity one and strictly larger in modulus than the rest of the eigenvalues~\cite{seneta:06}. We denote by $U$ and $V$ left and right eigenvectors of $\mathcal{L}_{\mathcal{H}_{\boldsymbol{\beta}}}$ corresponding to the eigenvalue $\rho$. Notice that $U_i>0$ and $V_i>0$, for all $i \in \mathcal{A}^{N}_R $. %When $\mathcal{H}_{\beta}$ is such that the multiplication of $\mathcal{L}_{\mathcal{H}_{\boldsymbol{\beta}}}$ with the vector of ones \textbf{1} is such that $\mathcal{L}_{\mathcal{H}_{\boldsymbol{\beta}}}\textbf{1} = \textbf{1}$, we say that the energy function $\mathcal{H}_{\beta}$ is normalized. In that case, if $q$ is a probability measure, $\cL^*_{\mathcal{H}_{\boldsymbol{\beta}}}(q)$ is also a probability measure.\\

The RPF matrix is not the Markov matrix we are looking for, moreover, is not a stochastic matrix, but can be converted into a stochastic matrix.  We recall that for an irreducible matrix  $M$ with spectral radius $\lambda$, and positive right eigenvector $\bf{v}$ associated to $\lambda$, then the {\em stochasticization} of $M$ is the following stochastic matrix:

\beq \label{stochasticization}
S(M)= \frac 1\lambda D^{-1}MD \ , 
\eeq
\noindent
where $D$ is the diagonal matrix with diagonal entries $D(i,i)=\bf{v}_{i}$. 
So, in our context, the MEMC transition matrix is given as follows:
\beq\label{pmar}
P=S(\mathcal{L}_{\mathcal{H}_{\boldsymbol{\beta}}}),
\eeq
whose unique stationary probability measure $\pi$ is explicitly given by 
\beq\label{ssmep}
\pi_i:=\frac{U_i\, V_i}{\langle U,V \rangle}, \quad \forall i \in \mathcal{A}^{N}_R,
\eeq
where $\langle U,V \rangle$ is the standard inner product in $\mathbb{R}^{NR}$ (we refer the reader to~\cite{seneta:06} for  details and proofs).

%Next, for further purposes \textcolor{red}{(me parece confuso introducir definiciones que no se van a usar hasta despues, o antes. Mejor introducir justo donde se usa, no?)}
%\textcolor{blue}{claro, yo puse eso porque tampoco sabia como acomodarlo, aunque creo que con la propuesta de cambiar el orden de las secciones queda en el lugar correcto (despues tambien se usa para linear response)}

%\textcolor{green}{Podrias ver esto Cesar?}, 
%\textcolor{blue}{sorry aun no acabo de modificar esto de aca, hoy durante el dia si que tengo que acabar...}

\subsubsection{Thermodynamic formalism} %\textcolor{green}{Cesar esto hay que eplicarlo nuevamente pues asi no queda claro. $\mathcal{P}[\mathcal{H}_{\beta}]$ es una constante para $\beta$ fijo y resuelve el constrained maximization problem cuando se satisface (17). Lo usual en formalismo termodinamico es presentar primero el problema variacional y despues Gibbs (que resuelve el problema variacional). Yo seguiria esa linea pues aca ho hay nada nuevo.}

In the previous section we have shown how to obtain the transition matrix and the invariant measure of a Markov chain. However, we have not yet included the constraints (we have just used the features to build the energy function), in other words, we have not yet fit the parameters of the MEMC. %In order to fit the maximum entropy parameters we need to introduce more tools.
In order to fit the maximum entropy parameters
let us introduce the following quantity, 
%the variational principle as the following unconstrained problem:

\beq\label{VarPrinc}
\p{\mathcal{H}_{\boldsymbol{\beta}}}=\sup_{q \in \cM_{st}} \Big\{\mathcal{S}\{q\}\, + \, \mathbb{E}_{q}\left\{\mathcal{H}_{\boldsymbol{\beta}}\right\} \Big\} %=
%\mathcal{S}\{p\}\, + \, \mathbb{E}_{p}\left\{\mathcal{H}_{\boldsymbol{\beta}}\right\}
\eeq
where $\cM_{st}$ is the set of all stationary probability measures in $\mathcal{A}_N^R$ and $\mathbb{E}_{q}\left\{\mathcal{H}_{\beta} \right\}= \sum_{k=1}^K \beta_k \, \mathbb{E}_{q}\left\{f_k \right\}$ is the average value of $\mathcal{H}_{\beta}$ with respect to $q$. Solving the optimization problem \eqref{VarPrinc} one gets the Markov measure we are looking for.
%$p$ is 
%one can uses problem:
 %For finite range potentials the solution to \eqref{VarPrinc} is unique and is a Gibbs measure in the sense of Bowen that we define below (for further details on the thermodynamic formalism see~\cite{bowen:98}, for instance).
%---
%For the case of finite range energy functions, the corresponding Gibbs measure in the sense of Bowen satisfies an optimization problem known in the thermodynamical formalism as \emph{variational principle} \cite{bowen:98}.
%---
Indeed, one knows from the thermodynamical formalism (see ~\cite{bowen:98}) that 
%Just as before, take 
for our energy function $\mathcal{H}_{\beta}$ of range $R\geq 2$, %then it is known \cite{bowen:98} that 
there exists an unique translation invariant (stationary) Markov measure $p$ associated to $\mathcal{H}_{\beta}$ for which one has the constant $M >1$ %and $\mathcal{P}[\mathcal{H}_{\beta}] \in \mathbb{R}$ 
such that,

\beq\label{gsb}
M^{-1} \leq \frac{p\{\boldsymbol{x}_{1,n}\}}{\exp(\sum_{k=1}^{n-R+1} \mathcal{H}(\boldsymbol{x}_{k,k+R-1}) -(n+R-1)\mathcal{P}[\mathcal{H}_{\beta}])}\leq M,
\eeq
\noindent
that attains the supremum \eqref{VarPrinc}, that is
$\mathcal{S}\{p\}\, + \, \mathbb{E}_{p}\left\{\mathcal{H}_{\boldsymbol{\beta}}\right\}$.
%Observe that the measure $p$ is prescribed by the energy function of spike blocks and their transitions, and the invariant measure is given by the eigenvectors of $\mathcal{L}_{\mathcal{H}_{\beta}}$. 
The quantity $\mathcal{P}[\mathcal{H}_{\beta}]$ is called \emph{topological pressure}, which plays the role of the free energy in the statistical mechanics. The measure $p$, as defined by~\eqref{gsb}, is known as the Gibbs measure in the sense of Bowen. Note that one can show that MEMCs are particular cases of  these measures, associated to finite-range energy functions. Moreover, \eqref{maxentpd} is a particular case of \eqref{gsb}, when $M=1$ and  $\mathcal{H}_\beta$ is an energy function of range one.

The average values of the features, their correlations, as well as their higher cumulants can be obtained by taking the successive derivatives of the topological pressure with respect to their conjugate parameters $\boldsymbol{\beta}$.
This explains the important role played by the topological pressure in this framework.
In general,
\beq\label{moyn}
\frac{\partial^n \p{\mathcal{H}_{\boldsymbol{\beta}}}}{\partial \beta_k^n} =\kappa_n \quad \forall k \in \{1,...,K\},
\eeq
\noindent
where $\kappa_n$ is the cumulant of order $n$ (see appendix \ref{Appen-cum-gen}.).
\noindent
In particular, taking the first derivative: 

\beq\label{moy}
\frac{\partial \p{\mathcal{H}_{\boldsymbol{\beta}}}}{\partial \beta_k} =\mathbb{E}_p\{f_k\}=c_k, \quad \forall k \in \{1,...,K\},
\eeq
where $\mathbb{E}_p\{f\}$ is the the average of $f_k$ with respect to $p$ (maximum entropy measure), which is equal (by assumption) to the average value of $f_k$ with respect to the empirical measure from the data $c_k$, that constraint of the maximization problem. These equations suggest a relationship with the logarithm of the free energy or log partition function of the Boltzmann Gibbs distribution. Indeed, for range one potentials (time-independent Maximum entropy distributions) $\rho (\boldsymbol{\beta})=Z (\boldsymbol{\beta})$ and  $\mathcal{P}[\mathcal{H}_{\boldsymbol{\beta}}] = \ln Z (\boldsymbol{\beta})$ which relates (\ref{fitr1}) with (\ref{moy}) (For a detailed example see section \ref{3rd}). This problem of estimating the MEMC parameters become computationally expensive for big matrices. However, there exist efficient algorithms to estimate the parameters for the Markov maximum entropy problem in the literature~\cite{nasser-marre-etal:14}. 

\begin{figure}[h!]
  \centering
\    \includegraphics[width=0.7\textwidth]{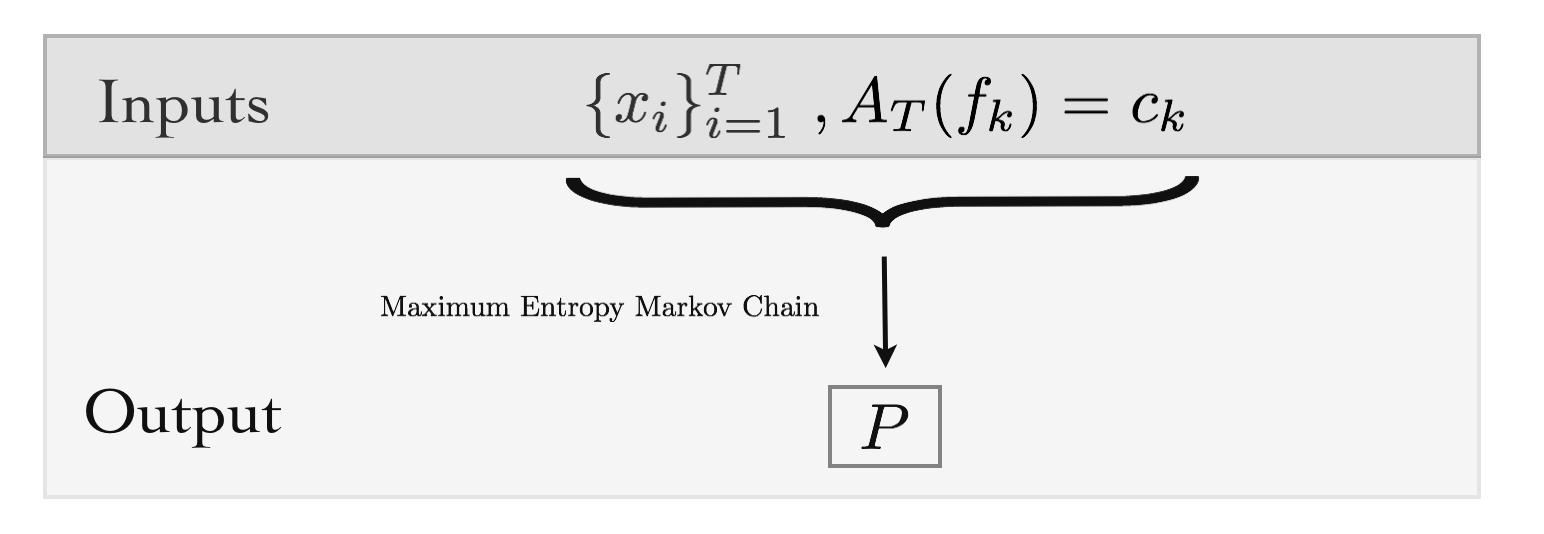}
    \caption{Algorithmic view of the MEMC: Inputs are the spike trains $\{x_i\}_{i=1}^T$ and the average values of a set of features. The output is the MEMC transition matrix $P$}
     \label{fig:fr}
\end{figure}

\section{Large deviations and applications in MEMC} \label{sec:LDandapp}

%\subsubsection{Convergence properties}
%\textcolor{red}{ESta seccion tiene una estructura complicada, y en mi opinion poco clara. Habria que explicar al principio para donde va la cosa, y cuales son los contenidos que se van a presentar en cada subseccion.}

\subsection{Preliminary considerations}

This subsection reviews two elementary tools for studying the convergence of random variables while providing corresponding references. In the sequel, first the central limit theorem is introduced in subsection~\ref{sec:CLT}, and then large deviation theory is discussed in subsection~\ref{sec:ld}.

\subsubsection{Central limit theorem}\label{sec:CLT}

Let us first assume that one can have access to arbitrarily large data sequences. Consider $t\in \mathbb{N}$ and let $x_{0,t-1}$ be the spike-block of length $t$ (which is allowed to increase), and let $f:\Omega \to\mathbb{R}$ be an arbitrary feature (not necessarily belonging to the set of features chosen to fit the MEMC). In this section we establish asymptotic properties of $A_t(f)$ sampled with respect to the MEMC characterized by $p\{\cdot\}$. 

Through this work, we will assume that $p\{\cdot\}$ is an ergodic Markov probability measure, this meaning that every spiking block in $\mathcal{A}_R^N$ is attainable from every other block in the Markov chain within $R$ time steps as discussed in section \ref{sec:3}. Thanks to the ergodic assumption, it is guaranteed that the empirical averages become statistically more accurate as the sampling size grows~\cite{levinAl:09}, i.e., 
%\textcolor{red}{(pregunta: es necesario hacer los argumentos de las fs en $A_T(f)$ no sobrelapados para que el teo valga, o da lo mismo si hay sobrelapes?)}\textcolor{blue}{esta es una buena pregunta, segun entiendo funciona con overlapping aunque es mas tricky las prueba y hay que considerearlas mas bien cuando estimas la entropia mediante los hitting times, aca no tenemos problema segun entiendo}
$$
A_t(f) \rightarrow \mathbb{E}_{p}\{f\}.
$$
However, the above result does not clarifies the rate at which the estimate accuracy improves. For answering this question, one can use the central limit theorem (CLT) for ergodic Markov chains (see ~\cite{jones:04}). This theorem states that there exists a constant $\sigma>0$ 
%(\textcolor{red}{mejor usar otra letra, esto parece mucho varianza pero no lo es...}\textcolor{blue}{en realidad en el caso de sistemas con correlaciones corresponde a la varianza, podriamos poner $\sqrt{Cov_{f}}$ donde $Cov_{f}$ es la funcion de autocovarianza, pero se vuelve cumbersome la notacion,no? quiza no tanto }) 
such that for large values of $t$, the random variable $\frac{\sqrt{t}}{\sigma}\big[ A_{t}(f) - \mathbb{E}\{f\}\big]$ distributes as a standard Gaussian random variable\footnote{Technically, the central limit theorem says that 
\[
p\left\{\frac{\sqrt{t}}{\sigma}\big[ A_{t}(f) - \mathbb{E}\{f\}\big]\leq x\right\}\to \frac{1}{\sqrt{2\pi}\sigma}\int_{-\infty}^{x}e^{-\frac{s^{2}}{2\sigma}}ds,
\]
where the convergence is in distribution.}, with $\sigma$ being the square-root of the auto-covariance function of $f$~\cite{jones:04}. This, in turn, implies that ``typical'' fluctuations of $A_t(f)$ around its mean value $\mathbb{E}\{f\}$ are of the order of $\sigma / \sqrt{t}$. %(c.f. Appendix \textcolor{green}{esto lo hago yo mañana}\textcolor{blue}{esta ahora en la sec 4.4.2}).

\subsubsection{Large deviations}~\label{sec:ld} 
%\textcolor{red}{Que se gana presentando LDT antes de la secion 4? Siendo que no es algo tan bien conocido por la gente no matematica, yo moveria 2.4.2 y 2.4.3 a la Secion 4 y presentaria todo de manera unificada. No creo que se gane mucho presentando esto primero aca, de manera un poco abstracta, separando los contenidos en dos partes. Yo mejor juntaria todo el contenido respecto a convergencia, y lo presentaria bien dirigido a la problematica especifica que se ve en ch4.}
Although the central limit theorem for ergodic Markov chains is accurate in describing typical events (which are fluctuations around the mean value), it does not say anything about the likelihood of larger fluctuations. Despite that it is clear that the probability of such large fluctuations goes to zero as the sample size increases, it is valuable to describe the corresponding decrease rate. In particular, one says that an empirical average $A_{t}(f)$ satisfies a large deviation principle (LDP) with rate function $I_{f}$, defined as
\beq\label{ldpa}
I_f(s) := -\lim_{t\rightarrow\infty} 
\frac{1}{t} \log p\{ A_{t}(f) > s\},
\eeq
if the above limit exists. Intuitively, the above condition for large $t$ implies that $p\{ A_t(f) >s\} \approx e^{-t I_f(s)}$. In particular, if $s > \mathbb{E}_p\{f\}$ the Law of Large Numbers (LLN) guarantees that $p\{ A_{t}(f) > s\}$ tends to zero as $t$ grows; the rate function quantifies the speed at which this happens. %probability goes to zero (exponential with respect to $s$). 

Calculating $I_{f}$ directly, i.e. by using the definition (eq \ref{ldpa}), can be a formidable task. However, the G\"artner-Ellis theorem provides a smart shortcut for avoiding this problem~\cite{ ellis:85}. To this end, let us introduce the \textit{scaled cumulant generating function} (SCGF)\footnote{The name comes from the fact that the $n$-th cumulant of $f$ can be obtained by $t$ successive differentiation operations over of $\lambda_f(k)$ with respect to $k$, and then evaluating the result at $k=0$.} associated to the random variable $f$, by
\beq\label{scgf}
\lambda_f(k)=: \lim_{t \rightarrow \infty} \frac{1}{t} \ln \mathbb{E}_p\left[e^{tkA_{t}(f)}\right], \quad k \in \mathbb{R}, 
\eeq
when the limit exists (further general details about cumulant generating functions are found in Appendix~\ref{Appen-cum-gen}). Note that, while $A_t(f)$ is an empirical average taken over a sample, the expectation in \eqref{scgf} is taken over the probability distribution given by the corresponding model $p\{\cdot\}$. If $\lambda_{f}$ is differentiable, then the G\"artner-Ellis theorem ensures that the average $A_{t}(f)$ satisfies a LDP with rate function given by the Legendre transform of $\lambda_{f}$, that is
%
%%The next theorem relates the SCGF and the large deviations rate function.\\
%The function $\mathbb{E}\left[e^{kX}\right]$ for k real is known as the generating function of  $X$; $ \ln \mathbb{E}\left[e^{kX}\right]$ is known as the log-generating function or cumulant generating function. The word 'scaled' comes from the extra factor $1/n$. 

%Consider the case where the sequence of random variables $X^0, . . . , X^n$ form a Markov chain i.e. the random variables $x^i$'s are time-dependent. This means that the joint pdf $p(x^0, . . . , x^n)$ has the form
%$$p(x^0, . . . , x^n)= \pi(x^0) \prod_{i=0}^{n-1} P(x^{i+1} \mid x^i);$$
%\\
%where $\pi(x^0)$ is some initial p.d.f for $x^0$ and $P(x^{i+1} \mid x^i)$ is the transition
%probability density that $x^{i+1}$ follows $x^i$ in the Markov sequence $x^0 \rightarrow x^1 %\rightarrow  \cdots \rightarrow  x^n$.

%$$\bar{f}_n = \frac{1}{n} \sum_{i=0}^{n-1}f(x^i,x^{i+1})$$

%\noindent
%\textbf{G\"{a}rtner-Ellis theorem: } If $\lambda_f(k)$ is differentiable, then there exist a large deviation principle for the average process $A_{T}(f)$ whose rate function is:
\beq\label{rfge}
I_f(s)= \max_{k \in \mathbb{R}} \{ks - \lambda_f(k)\}. 
\eeq
\noindent
Therefore, in summary, one can study the large deviations of empirical averages $A_t(f)$ by first computing their SCGF from the selected model and then finding their Legendere transform.

%\subsubsection{Connections with LLN and CLT}

One of the most useful applications of the LDP is to estimate the likelihood that $A_t(f)$ adopts a value far from its expected value. For illustrate this, let us assume that $I_f(s)$ is a positive differentiable convex function\footnote{A classical result in LDP states that $I_f(s)$ is a convex function if $\lambda_f(k)$ is differentiable \cite{dembozeitouni:10}. For a discussion about the differentiability of  $\lambda_f(k)$ see \cite{touchette:12}.}. Then, because of the properties of convex functions $I_f(s)$ has a unique global minimum. Denoting this minimum by $s^*$, it follows from the differentiability of $I_f(s)$ that $I_f(s^*)=0$, and using properties of the Legendre transform $s^*=\lambda'_f(0)=\lim_{t \rightarrow \infty} \mathbb{E}_p(f)$. This is the LLN, i.e., $A_t(f)$ gets concentrated around $s^*$. Consider a value $s \neq s^*$ and assume that $I_f(s)$ admits a Taylor series around $s^*$ given by
$$I_f(s) = I_f(s^*) + I'_f(s^*)(s-s^*) +\frac{I''_f(s^*)(s-s^*)^2}{2}+ O(s-s^*)^3$$
\noindent%
Since $s^*$ must correspond to a zero and a minimum of $I(s)$ , the first two terms in this series vanish, and as $I(s)$ is convex function $I''(s)>0$. For large values of $k$, we obtain from (\ref{ldpa})

\begin{equation} \label{eq1}
\begin{split}
p\{ A_t(f) > s\} & \approx e^{-tI_f(s)} \\
 & \approx e^{-t\left(\frac{I''_f(s^*)(s-s^*)^2}{2}\right)}
\end{split}
\end{equation}

\noindent
so the small deviations of $A_t(f)$ around $s^*$ are Gaussian-distributed as for i.i.d. sums $1/I''_f(s^*)=\lambda''_f(0)=\sigma^2$. In this sense, large deviation theory can be seen as an extension of the CLT because it gives information not only about the small deviations around $s^*$ but also about large deviations (not Gaussian) of $A_t(f)$.

\subsection{Large deviations for features of MEMC}\label{LD} 

%\textcolor{red}{Esto sale un poco out of the blue. Que se quiere hacer? Mejor explicar para donde va la cosa antes de definir cosas sin saber para donde se va...}

In this section, we focus on the statistical properties of features sampled from the inferred MEMC. For example, one may be interested in measuring the probability of obtaining "rare" average values of features like firing rates, pairwise correlations, triplets or spatiotemporal events. This characterization is relevant as these features are likely to play an important role in neuronal information processing, and rare values may hinder the whole enterprise of conveying information. We show in this section how to obtain the large deviations rate functions of arbitrary features through the G\"{a}rtner-Ellis theorem via the SCGF. In particular, we show that the SCGF can be directly obtained from the inferred Markov transition matrix $P$. 

%%In order to obtain the SCGF associated with our random variables $A_{t}(f)$ one introduces the tilted transition matrix. So that, it ables us to get the rate function from the data, just as describe in figure~\ref{fig:f1}.

Consider MEMC taking values on the state space $\mathcal{A}^{N}_R$ with transition matrix $P$. Let $f$ be a feature of range $R$ which consider only the block  and $k\in\mathbb{R}$, we introduce $\widetilde{P}^{(f)}(k)$,  the \textit{tilted transition matrix by f} of $P$, parametrized by $k$, whose elements are given by:

\beq\label{tilt}
\widetilde{P}^{(f)}_{ij}(k)=P_{ij}e^{kf(j)} \quad i,j \in \mathcal{A}^{N}_R.
\eeq
\noindent

%In the case when $f$ is a two time step feature the tilted transition matrix is defined as follows:
%\beq\label{tilt2}
%\widetilde{P}^{(f)}_{ij}(k)=P_{ij}e^{kf(i,j)}.
%\eeq
%Although this can be generalized for $n$ time step features we will restrict our selves to one and two time steps hereafter. 

For transition matrices $P$ inferred from the MEP, the tilted transition matrix can be built directly from the spectral properties of the transfer matrix (\ref{transfmatrix}) as follows, 

\begin{eqnarray}\label{tme1}
\widetilde{P}^{(f)}_{ij}(k) &=& \frac{e^{\mathcal{H}_{\beta}(i,j)} V_{j}}{V_{i} \, \rho} e^{kf(j)}\\ \nonumber
  &=&  \frac{e^{\left[\mathcal{H}_{\beta}(i,j)+kf(j)\right]}V_{j}}{V_{i} \, \rho}  \quad i,j \in \mathcal{A}^{N}_R.
\end{eqnarray} 
\noindent
Recall that $V$ is the right eigenvector of the transfer matrix $\mathcal{L}$. Here we also have used the shortcut notation $\mathcal{H}_{\beta}(i,j)$ to indicate that the energy function takes the contributions from the blocks $i$ and $j$. Remarkably, the feature $f$ does not need to belong to the set of chosen features to fit the MEMC. 

Now, we can take advantage of the structure of the given process in order to obtain more explicit expressions for the SCGF $\lambda_{f}(k)$, for instance, if one considers i.i.d. random variables $X$ then, from the very definition one can obtain that

\[
\lambda(k)= \lim_{t \rightarrow \infty} \frac{1}{t}  \ln\mathbb{E}[e^{kX}]^t =\ln  \mathbb{E}[e^{kX}],
\]
which is the case of range one features. So, using equation \eqref{tilt}, we get that the maximum eigenvalue of the tilted matrix, denoted by $\rho(\widetilde{P}_{f}(k))$ is, 

\[
\rho \big(\widetilde{P}_f(k)\big)=  \sum_{j} \pi_j e^{kf(j)} \quad j \in \mathcal{A}_1^N.
\]
\noindent
Since $\widetilde{P}_{f}$ is a positive matrix the Perron-Frobenius theorem ensures the uniqueness of $\rho$ .

\begin{figure}[h!]
  \centering
\    \includegraphics[width=0.7\textwidth]{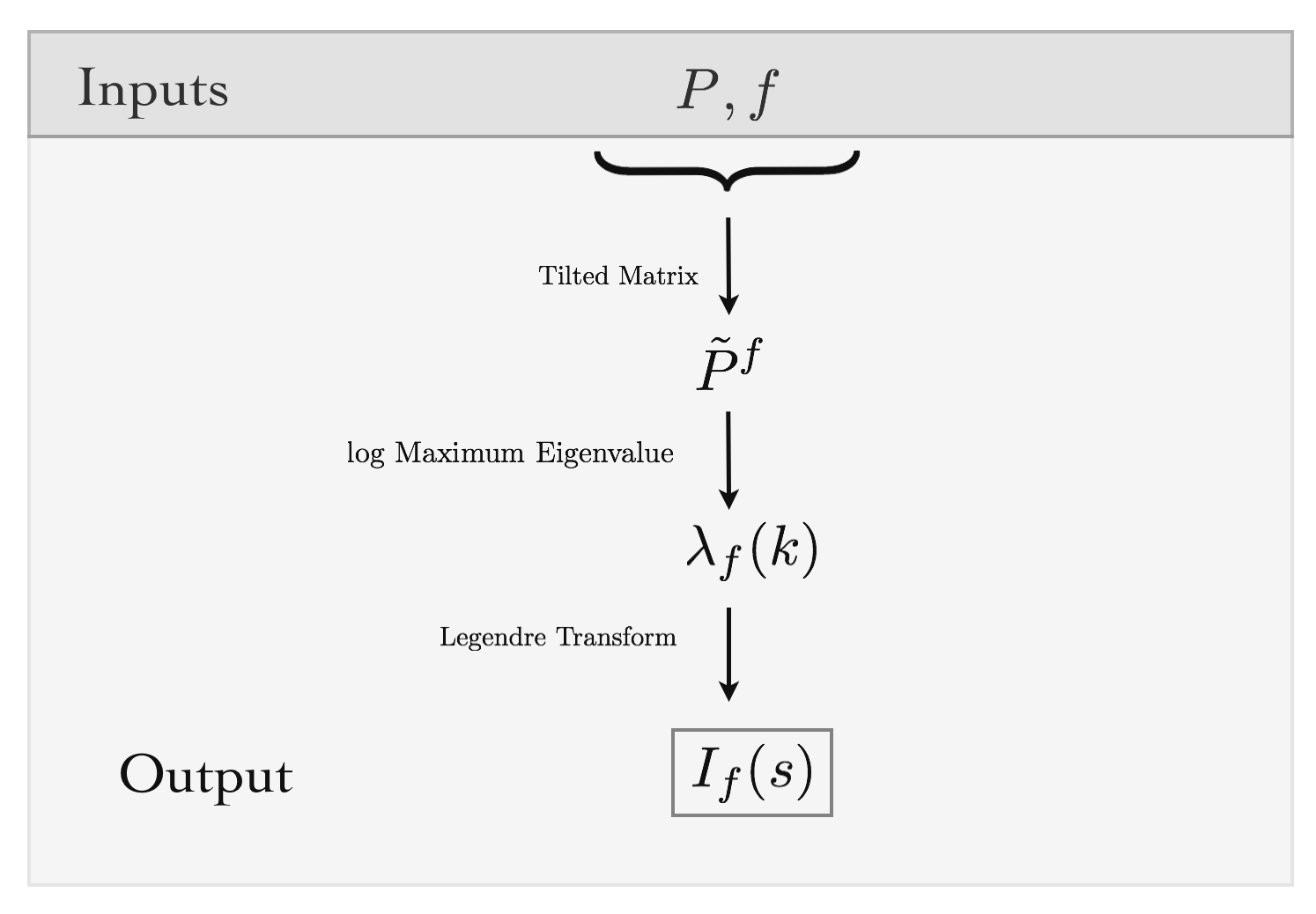}
    \caption{Algorithmic view of the method: Inputs are the maximum entropy Markov transition matrix and a feature. From the inputs the tilted transition matrix can be built. The rate function of the feature is obtained as the Legendre transform of the log maximum eigenvalue of the tilted transition matrix. Using this framework we can estimate the large deviations of the average values of the features.}
     \label{fig:f1}
\end{figure}

Next, for the case of additive features, one deals with positive Markov chains, and under the assumption of ergodicity, an straightforward calculation (see for instance~\cite{ellis:10}) leads us to obtain that
\begin{equation}\label{tev}
\lambda_f(k)= \ln \big(\rho \big(\widetilde{P}^{(f)}\big)\big).
\end{equation}
\noindent
It also can be proved  that $\lambda_f(k)$, in this case is  differentiable~\cite{ellis:10}, setting up the scene to apply the G\"{a}rtner-Ellis theorem,  which bypasses the direct calculation of $p\{ A_T(f) >s\}$ in \eqref{ldpa}, i.e., having $\lambda_f(k)$, its Legendre transform leads to the rate function of $f$ as shown in figure \ref{fig:f1}.

\subsection{Large deviations for the entropy production}%\label{varpro}

A stochastic process is said to be in equilibrium if one cannot notice the effect of time on it. It is worth noticing that non-equilibrium stochastic processes are natural candidates to model spike train statistics as time plays a non-symmetrical role \cite{cofre-mal:18}. One of the consequences of including features of range $R>1$ as constraints in the maximum entropy problem is that it opens the possibility to break the time-reversal symmetry present in the time-independent models. This captures the irreversible character of the underlying biological process and thus, allows to fit more realistic statistical models from the biological point of view.

To characterize this mathematically, we study how the distribution $p\{\cdot\}$ changes when the time order is reversed.
For this aim, let us consider a spike block $\boldsymbol{x}_{0,T-1} = \boldsymbol{x}_0, \boldsymbol{x}_1, \dots , x_{T-1}$ containing $T$ spike patterns, and define the time-reversed spike block $\boldsymbol{x}_{0,T-1}^{(R)}$ obtained by re-ordering the time index in reverse order, i.e., $\boldsymbol{x}_{0,T}^{(R)} =\boldsymbol{x}_{T-1}, \boldsymbol{x}_{T-2}, \dots , \boldsymbol{x}_2, \boldsymbol{x}_0$. 

A spiking network modeled by $p\{\cdot\}$ is said to be in equilibrium if $p\{\boldsymbol{x}_{0,T}\} = p\{\boldsymbol{x}_{0,T}^{(R)}\}$ for all $\boldsymbol{x}$ \cite{maes:99}. For a homogeneous discrete time ergodic Markov chain characterized by the Markov measure $p(\pi,P)$ taking values in $\mathcal{A}^{N}_R$, to be in equilibrium is equivalent to satisfy the \textit{detailed balance conditions}, which is given by the following set of equalities: 
$$
\pi_i P_{ij}=\pi_j P_{ji}, \quad \forall i,j \in \mathcal{A}^{N}_R.
$$
\noindent
Conversely, when these conditions are not satisfied the statistical model of the spiking activity is said to be a non-equilibrium system. 
Since non-equilibrium is expected to occur generically in neuronal network models, one would like to quantify how far from equilibrium is the inferred MEMC. For this purpose one can define the \textit{information entropy production} (IEP) for $p$, which is given by
\[
IEP(p):=\lim_{t\to \infty}\frac{1}{t}\ln\Bigg[\frac{p\{\boldsymbol{x}_{0,t-1}\}}{p\{\boldsymbol{x}_{0,t-1}^{(R)}\}}\Bigg],
\]
when the limit exists. For the maximum entropy Markov measure $p(\pi, P)$, the IEP is explicitly given by: 
\beq\label{iepde}
IEP(\pi, P)=\frac{1}{2}\sum_{i,j \in \mathcal{A}^{N}_R}\big[ \pi_{i}P_{ij}-  \pi_{j}P_{ji}\big]\log\frac{\pi_{i}P_{ij}}{\pi_{j}P_{ji}},
\eeq
(see~\cite{gaspard:04} for the calculation). We  remark that it is still possible to obtain the information entropy production rate also in the non-stationary case. Clearly, for features of range one, $IEP=0$ always, meaning that the process is  time-reversible, therefore the probabilities of every path and its corresponding time-reversal path are equal. For features of range $R>1$, $IEP>0$ generically (we refer the interested reader to~\cite{cofre-mal:18} for details and examples).

However, since in practice one only have access to limited amount of data, a natural question is to ask for the entropy production of the system considered up to a finite amount of time. It turns out that this characterization can be obtained through a LDP. With this in mind one may define the following feature:
\[
W_T(\boldsymbol{x}_{0,T-1})=\frac{1}{T}\ln\Bigg[\frac{p(\boldsymbol{x}_{0,T-1})}{p(\boldsymbol{x}_{0,T-1}^{(R)})}\Bigg].
\]
Since we have assumed that samples are produced by a stationary ergodic Markov chain characterized by $p(\pi, P)$, the ergodic theorem assures that for $p$-almost every sample, the quantity $W_{t}$ when $t$ goes to infinity converges, and it is by definition the IEP, 
\[
\lim_{t \rightarrow \infty} W_{t}(\boldsymbol{x}_{0,t-1}))=IEP(\pi,P).
\]
Once we have the convergence for $W_{t}$, we may ask for its large deviation properties. Under the same idea above, and following~\cite{jiang:04},  we introduce the following matrix:
%It can be shown  that defining the following matrix:
\[
F_{ij}=P_{ij} \ln \Bigg[\frac{\pi_i P_{ij}}{\pi_j P_{ji}}\Bigg]^k \quad i,j \in \mathcal{A}^{N}_R,
\]
this matrix help us to build the SCGF associated to $W_{t}$, through the logarithm of the maximum eigenvalue $\rho_F(k)$. Using the G\"{a}rtner-Ellis theorem one gets the rate function $I_W(s)$ for the IEP.

%\noindent
%The SCGF associated to $W_t$ denoted by $\lambda_W(k)$ can be found as the logarithm of the maximum eigenvalue $\rho_F(k)$ of the following matrix of positive elements \cite{jiang:04}: %\cite{quian:04}:

%\[
%\widetilde{P}_{ij}^{(W)}(k)=P_{ij}e^{kF_{ij}}.
%\]
%\noindent
%From the unique maximum eigenvalue we obtain the SCGF of $W$\cite{quian:04}:

%$$
% \lambda_W(k)=\lim_{t \rightarrow \infty} \frac{1}{t} \ln \mathbb{E}\left[e^{tkW_t}\right]=\ln \rho_F(k)$$

%$$
%I_W(s)= \max_k \{ks - \lambda_W(k)\} 
%$$

%\subsection{Large deviations to distinguish MEMC} 
\subsection{Large deviations and MEMC distinguishability} 

It is clear that there exist a relationship between accuracy of the estimation and sample size. The larger the sample size the more information is available and the uncertainty diminish. In the context of maximum entropy models, this idea has been well conceptualized using tools from information geometry \cite{amari:10,balasubramanian:97}. The main idea of this approach is that the maximum entropy models form a manifold of probability measures whose coordinates are the parameters $\boldsymbol{\beta}$. Consider a spike train dataset $\boldsymbol{x}_{0,T-1}$ consisting of $T$ spike patterns obtained from a spiking neuronal network. Given a set of features $\{f_{k}\}_{k=1}^K$ and their empirical averages, one may infer the parameters $\boldsymbol{\beta}=(\beta_1,\dots,\beta_K)$ characterizing the MEMC $p(\pi, P)$. We may use the inferred MEMC to generate a sample $\boldsymbol{x'}_{0,T-1}$ of the same size as the original dataset. Considering the same set of features one may apply again the MEP to infer a new set of parameters $\boldsymbol{\beta'}$ from  $\boldsymbol{x'}_{0,T-1}$, which is, in general, expected to be different from $\boldsymbol{\beta}$. These maximum entropy models will belong to a certain volume in the manifold which will decrease as the sample size increase \cite{balasubramanian:97}. On the other hand, increasing the sample size of $\boldsymbol{x'}_{0,T-1}$, one expects that the Markov chain $p'(\pi', P')$ specified by $\boldsymbol{\beta'}$ gets "closer" to the one characterized by $\boldsymbol{\beta}$. The idea of relating a distance in the parameter space with a distance in the space of probability measures can be rigorously formulated using large deviations techniques. Let us start by defining the relative entropy between these two MEMC (Gibbs measures in the sense of Bowen \eqref{gsb}), which provides a notion of "distance" \footnote{The relative entropy is not a metric (is not symmetric nor satisfy the triangle inequality).}. In order to do that in the context of MEMC's consider a Gibbs measure $p$  associated to the energy function
$\mathcal{H}_{\boldsymbol{\beta}}$, and let $q$ be another Gibbs measure. The Ruelle-F\"{o}llmer theorem gives us an expression for the relative entropy density between two Gibbs measures in terms of the pressure, the entropy rate and the expected value of the energy function with respect to $q$ (see~\cite{georgii:03}), as follows:
\begin{equation} \label{rf}
\textsf{d}(q \mid p) = \mathcal{P}[\mathcal{H}_{\beta}] -S(q) -\mathbb{E}_{q}(\mathcal{H}_{\beta}).
\end{equation}
Observe that if $\textsf{d}(q \mid p)=0$, we obtain the variational characterization of Gibbs measures \eqref{VarPrinc}.

Consider the potential $\mathcal{H}_{\boldsymbol{\beta}}=\sum_{k=1}^K \beta_k f_k$ associated with a MEMC $p(\pi, P)$. Given a set of empirical averages $A_t(f_k)$ generated by a sample of $p(\pi, P)$ we obtain new maximum entropy parameters $\boldsymbol{\beta'}$. The probability that the maximum entropy parameters $\boldsymbol{\beta'}$ associated with an ergodic Markov Chain $p'(\pi', P')$ get close to $\boldsymbol{\beta}$ follow the following large deviation principle \cite{dembozeitouni:10}:

\begin{equation} \label{eq48}
\lim_{\delta \rightarrow 0} \lim_{t \rightarrow \infty} \frac{-1}{t} \ln \mathbb{P}\Big(\mid \boldsymbol{\beta} - \boldsymbol{\beta'}  \mid \in \Delta \delta  \Big) = \textsf{d}(p \mid p') ,
\end{equation}

%\textcolor{blue}{$|\beta-\beta'|$ es el vector distancia? o solo la diferencia, ya que $\Delta\delta \in[-\delta,\delta]^{K}$}

\noindent
where $\Delta \delta=[-\delta,\delta]^K$. Calling and the vector $\delta \boldsymbol{\beta}= \boldsymbol{\beta}-\boldsymbol{\beta'}$ and choosing $\Delta \delta$ close to 0 we informally rewrite the above corollary in the form:

\begin{equation} \label{eq8}
\frac{-1}{t} \ln \mathbb{P}\Big(\mid \delta \boldsymbol{\beta}  \mid \in \Delta \delta \Big) \underset{t \rightarrow \infty}{\longrightarrow} \textsf{d}(p \mid p') .
\end{equation}
\noindent
Thus, for large $T$ we get: 

$$\mathbb{P}\Big(\mid \delta \boldsymbol{\beta} \mid \in \Delta \delta \Big) \approx e^{-t  \textsf{d}(p \mid p')},$$
\noindent
which implies that close-by parameters are associated to close-by probability measures \cite{balasubramanian:97}.%\\

Consider now two MEMC $p(\pi, P)$ and  $p'(\pi', P')$ specified by $\mathcal{H}_{\boldsymbol{\beta}}$ and $\mathcal{H}_{\boldsymbol{\beta'}}$ respectively with the same family of features. We say that the MEMC's are $\epsilon$-\textit{indistinguishable} if: 

\begin{equation} \label{eq9}
- \ln \mathbb{P}\Big(\mid \delta \boldsymbol{\beta}  \mid \in \Delta \delta \Big) \leq \epsilon .
\end{equation}
%\textcolor{blue}{aqui la definicion de indistinguibilidad me queda un poco rara, que no siempre es cero que $\beta =\beta'$? tambien estaria bien una cita para la definicion}
\noindent
As both MEMC's satisfy the variational principle (\ref{VarPrinc}), the relative entropy between $p$ and $p'$ (\ref{rf}) reads:

%\begin{equation} \label{eqps}
%\begin{split}
%\mathcal{P}[\mathcal{H}_{\beta}]& =S(\mu)+\mu(\mathcal{H}_{\beta}) \\
%\mathcal{P}[\mathcal{H}_{\beta'}] & =S(\mu')+\mu'(\mathcal{H}_{\beta'})
%\end{split}
%\end{equation}

\begin{equation} \label{rf2}
\textsf{d}(p \mid p')=\mathcal{P}[\mathcal{H}_{\beta'}]-\mathcal{P}[\mathcal{H}_{\beta}]+p(\mathcal{H}_{\beta})-p(\mathcal{H}_{\beta'}).
\end{equation}

\noindent
Taking the Taylor expansion of $\textsf{d}(p \mid p')$ around $\boldsymbol{\beta'}=\boldsymbol{\beta}$ we get:

$$\textsf{d}(p \mid p') \approx \textsf{d}(p \mid p) + \sum_{k}\frac{\partial \textsf{d}(p \mid p')}{\partial \beta'_k}\Bigr|_{\substack{\beta'=\beta}} (\beta_k-\beta_k')+ \frac{1}{2}\sum_{k,j}\frac{\partial^2 \textsf{d}(p \mid p')}{\partial \beta'_k \beta'_j}\Bigr|_{\substack{\beta'=\beta}} (\beta_k-\beta_k')(\beta_j-\beta_j').$$
\noindent
Since $\textsf{d}(p \mid p')$ is minimized at $\beta'=\beta$ we obtain,

$$\textsf{d}(p \mid p') \approx  \frac{1}{2}\sum_{k,j}\frac{\partial^2 \textsf{d}(p \mid p')}{\partial \beta'_k \beta'_j}\Bigr|_{\substack{\beta'=\beta}} (\beta_k-\beta_k')(\beta_j-\beta_j').$$
\noindent
Taking the second derivative of $\textsf{d}(p \mid p')$ from (\ref{rf2}), one also has that,

\begin{equation} \label{matl}
\frac{\partial^2 \textsf{d}(p \mid p')}{\partial \beta'_k \beta'_j}=\frac{\partial^2 \mathcal{P}[\mathcal{H}_{\beta'}] }{\partial \beta'_k \beta'_j}=L_{kj}.
\end{equation}

\noindent
The second partial derivatives of the topological pressure with respect to the parameters $\beta'_k$ and $\beta'_j$ can be conveniently arranged in a matrix $L$ with components $L_{kj}$. Given two MEMC's specified by $\mathcal{H}_{\boldsymbol{\beta}}$ and $\mathcal{H}_{\boldsymbol{\beta'}}$, in the limit of large $t$ they are $\epsilon$-indistinguishable if:

\begin{equation} \label{eqet}
\frac{1}{2} \Big[( \delta \boldsymbol{\beta})^{\textsf{T}} L (\delta \boldsymbol{\beta})\Big] \leq \frac{\epsilon}{T} ,
\end{equation}

\noindent
where $\textsf{T}$ denotes transpose. The matrix $L$ can be obtained from data without need to fit the parameters. Equation (\ref{eqet}) characterize a region in the space of MEMC of indistinguishable models, whose volume can be calculated in the large $t$ limit using spectral properties of the matrix $L$ \cite{balasubramanian:97}.  This result generalizes a previous result for maximum entropy distributions for range one energy functions in \cite{mastromatteo:11}.\\

\section{Illustrative examples}

%\textcolor{green}{Si se les ocurre algun ejemplo o si podrian darme ideas de que poner en aca seria genial, ando falto de imaginacion}
%\textcolor{red}{Que tal un ejemplo simetrico sin produccion de entropia, versus otro de tres neuronas donde haya circulacion y por lo tanto produccion de entropia (aunque este en estado estacionario...)}
%\textcolor{blue}{eso es algo que yo me he preguntado, si es que hay algun ejemplito real donde haya una terna de neuronas que se conozca bien que tengan un tipo de correlacion temporal y que por lo tanto se le pueda aplicar el modelo de cadena de markov, etc y medir alguna prod. de entropia...}

In this section we illustrate the presented methods in some simple scenarios. In these examples we follow a set of steps:

\begin{enumerate}
\item Choose the observables and build the energy function (\ref{energy}).
\item Build the transfer matrix (\ref{transfmatrix}).
\item Compute the free energy and find the maximum entropy parameters using (\ref{moy}).
\item Build the Markov transition matrix using (\ref{pmar}).
\item Choose the observable to examine and build the tilted transition matrix using (\ref{tilt}).
\item Compute the Legendre transform of the $\log$ maximum eigenvalue of the tilted transition matrix to obtain the rate function (\ref{tev}).
\end{enumerate}

For the sake of clarity, in this section we focus on small neuronal networks. It is clear, however, that the extension of these techniques to larger neural populations is straightforward.

\subsection{First example: Maximum entropy model of a range 2 feature}

Consider spiking data from two interacting neurons. We measure only the average value of a of a range 2  feature from the spiking data to fit a MEMC. The feature denoted by $f_1$ is given by $\tilde{f}_1(\boldsymbol{x}_{0,1})=x_0^2 \cdot x_1^1$, which detects when a spike of the second neuron is followed by a spike in the first one. The system can be described with the help of an energy function $\mathcal{H}(\boldsymbol{x}_{0,1})=\beta_1 \tilde{f}_1(\boldsymbol{x}_{0,1})$.

For a given dataset of $T$ spike blocks of range 2 the empirical average reads,

\begin{equation} \label{eqr}
A_T(f_1) =c
\end{equation}
\noindent
this means that in the data one finds that this event appears $c\%$ of the time.

The transfer matrix $\mathcal{L}_{\mathcal{H}}$ (c.f. \eqref{transfmatrix}) associated with this energy function is a matrix indexed by the 16 states of the system, which in this case is the set $\mathcal{A}_2^2$: 

$$\left\lbrace \pare{\begin{array}{ccc}
0 & 0\\
0 & 0\\
\end{array}},\pare{\begin{array}{ccc}
0 & 0\\
0 & 1\\
\end{array}}, \dots,\pare{\begin{array}{ccc}
1 & 1\\
1 & 1\\
\end{array}} \right\rbrace.$$

%$$
%\cL_{\mathcal{H}}^*=
%\pare{
%\begin{array}{ccccccccccccccccccccccc}
%1 & 1 & 1 & 1 & 0 & 0 & 0 & 0 & 0 & 0 & 0 & 0 & 0 & 0 & 0 & 0  \\
%0 & 0 & 0 & 0 & 1 & 1 & 1 & 1 & 0 & 0 & 0 & 0 & 0 & 0 & 0 & 0  \\
%0 & 0 & 0 & 0 & 0 & 0 & 0 & 0 & 1 & e^{\beta_1} & 1 & e^{\beta_1} & 0 & 0 & 0 & 0  \\
%0 & 0 & 0 & 0 & 0 & 0 & 0 & 0 & 0 & 0 & 0 & 0 & 1 & e^{\beta_1} & 1 & e^{\beta_1}   \\
%1 & 1 & 1 & 1 & 0 & 0 & 0 & 0 & 0 & 0 & 0 & 0 & 0 & 0 & 0 & 0  \\
%0 & 0 & 0 & 0 & 1 & 1 & 1 & 1 & 0 & 0 & 0 & 0 & 0 & 0 & 0 & 0  \\
%0 & 0 & 0 & 0 & 0 & 0 & 0 & 0 & 1 & e^{\beta_1} & 1 & e^{\beta_1} & 0 & 0 & 0 & 0  \\
%0 & 0 & 0 & 0 & 0 & 0 & 0 & 0 & 0 & 0 & 0 & 0 & 1 & e^{\beta_1} & 1 & e^{\beta_1}   \\
%1 & 1 & 1 & 1 & 0 & 0 & 0 & 0 & 0 & 0 & 0 & 0 & 0 & 0 & 0 & 0  \\
%0 & 0 & 0 & 0 & 1 & 1 & 1 & 1 & 0 & 0 & 0 & 0 & 0 & 0 & 0 & 0  \\
%0 & 0 & 0 & 0 & 0 & 0 & 0 & 0 & 1 & e^{\beta_1} & 1 & e^{\beta_1} & 0 & 0 & 0 & 0  \\
%0 & 0 & 0 & 0 & 0 & 0 & 0 & 0 & 0 & 0 & 0 & 0 & 1 & e^{\beta_1} & 1 & e^{\beta_1}   \\
%1 & 1 & 1 & 1 & 0 & 0 & 0 & 0 & 0 & 0 & 0 & 0 & 0 & 0 & 0 & 0  \\
%0 & 0 & 0 & 0 & 1 & 1 & 1 & 1 & 0 & 0 & 0 & 0 & 0 & 0 & 0 & 0  \\
%0 & 0 & 0 & 0 & 0 & 0 & 0 & 0 & 1 & e^{\beta_1} & 1 & e^{\beta_1} & 0 & 0 & 0 & 0  \\
%0 & 0 & 0 & 0 & 0 & 0 & 0 & 0 & 0 & 0 & 0 & 0 & 1 & e^{\beta_1} & 1 & e^{\beta_1}   
%\end{array}
%}.
%$$
\noindent
As $\mathcal{L}_{\mathcal{H}}$ is primitive by construction (c.f. \eqref{transfmatrix}), it satisfies the hypothesis of the Perron-Frobenius theorem. In fact, its unique maximum eigenvalue is $\rho(\beta_1)=e^{\beta_1}+3$. Given the restriction \eqref{eqr}, using \eqref{moy} we obtain the following relationship between the parameter $\beta_1$ and the value of the restriction $c$:
$$ \frac{\partial \p{\mathcal{H}}}{\partial \beta_1} =\frac{\partial \log(e^{\beta_1}+3)}{\partial \beta_1} = \frac{e^{\beta_1}}{e^{\beta_1}+3} = c.$$
\noindent
This equation can be solved numerically. Using the obtained value of $\beta_1$ in equation \eqref{pmar} one can find the corresponding Markov transition matrix. Note that, among all the Markov chains that match exactly the restriction, the selected one maximizes the KSE. Moreover, it is direct to check that the variational principle \eqref{VarPrinc} is satisfied. Examples of values of $\beta_1$ for different values of $c$ and IEP (\ref{iepde}) for each value of $\beta_1$ are given in the following table: 

\begin{table}[h!]
\centering
\caption{ }
\label{my-label}
\begin{tabular}{| l | l |l |}
 \hline
 $ c $ & $\beta_1$ & $IEP$  \\ \hline
  0.043 & -2  & 0.176\\ \hline
   0.11 & -1  & 0.056\\ \hline
    0.25 & 0 & 0\\ \hline
   0.475 & 1 & 0.0525 \\ \hline
0.711 & 2 & 0.1184\\ \hline
\end{tabular}
\end{table}

\begin{figure}[h!]
  \centering
    \includegraphics[width=0.9\textwidth]{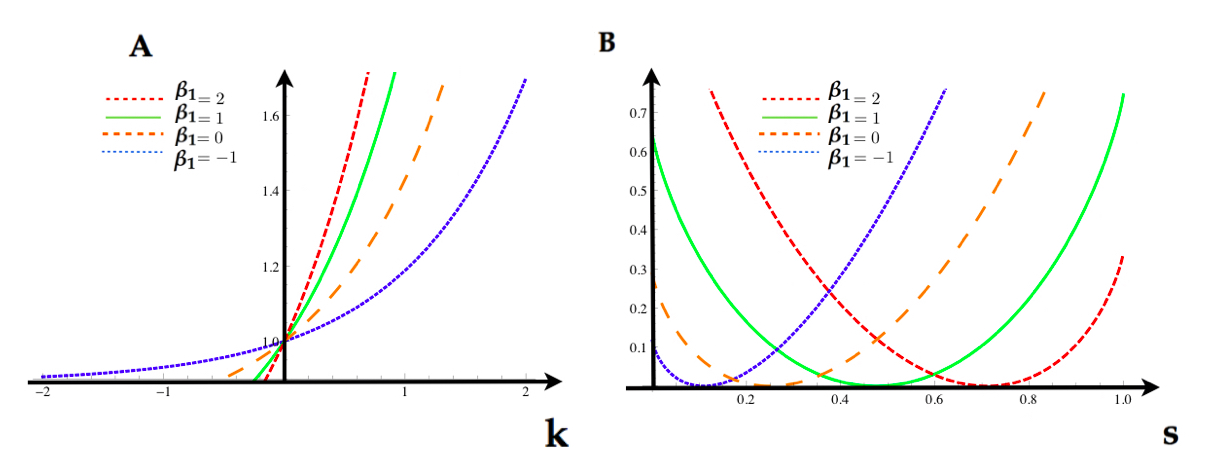}
    \caption{\textbf{A)} SCGF (\ref{tev}) for the feature $f_1$ of the first example computed at the values provided by the table above. \textbf{B)} Rate function for the same feature computed at the same parameter values as the SCGF. Each of this functions are obtained taking the Lagrange transform of the respective SCGF in \textbf{A)}. }
    \label{fig:tcr}
\end{figure}

\begin{figure}[h!]
  \centering
    \includegraphics[width=0.9\textwidth]{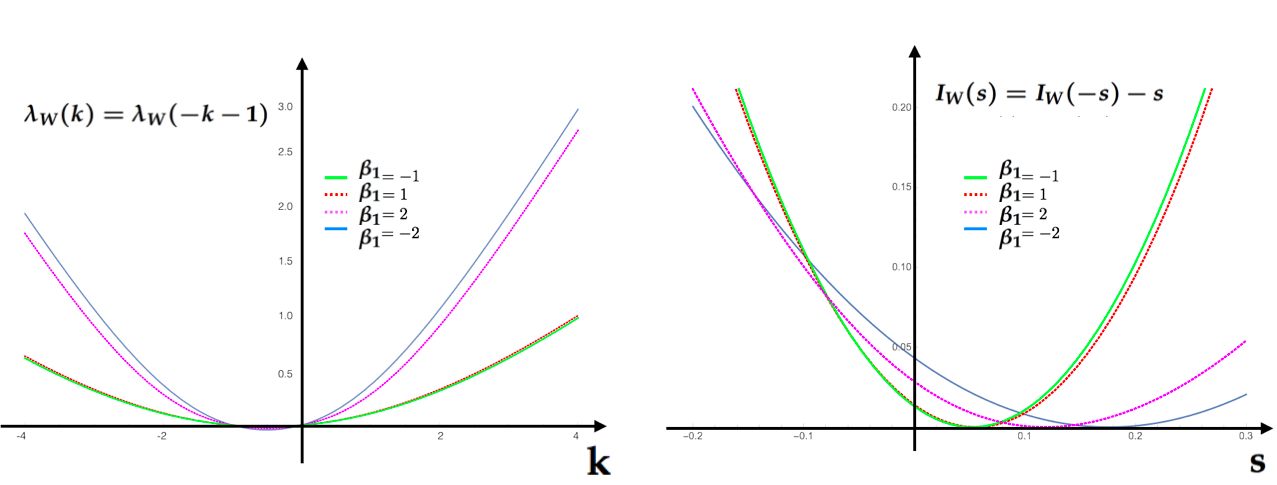}
    \caption{Gallavotti-Cohen symmetry relationship for the IEP for values in table 1.  Left SCGF $\lambda_W(k)$. Right rate function of the IEP feature $W, I_W(s)$.}
     \label{fig:gce}
\end{figure}

\noindent
Having the MEMC, we are now interested in analyzing the statistical fluctuations of the feature $f_1$. Using equation (\ref{tilt}) we obtain the tilted transition matrix $\widetilde{P}^{(f_1)}_{ij}(k)$ for each of the values in the table 1. In figure (\ref{fig:tcr}), we compute for each value of $\beta_1$ we compute the SCGF $\lambda_{f_1}(k)$ (\ref{tev}) and the Legendre transform (rate function) associated to the feature $I_{f_1}(s)$.
\noindent
In figure (\ref{fig:gce}), we compute for each value of IEP in the table the rate function and illustrate for this example the symmetry relationship (\ref{gcsr}).

\subsection{Maximum entropy model with only synchronous constrains}\label{3rd}

Let us now consider a network of three neurons. We focus here on range one features. In this example we consider features related to the firing rates and synchronous pairwise correlations (Ising model \cite{schneidman-berry-etal:06,tkacik-marre-etal:13}). Specifically, we consider the following energy function:
$$
\mathcal{H}(\boldsymbol{x}_0)=\beta_1 x_0^1 +\beta_2 x_0^2+\beta_3 x_0^3+\beta_4 x_0^1 \cdot x_0^2+\beta_5 x_0^1 \cdot x_0^3+\beta_6 x_0^2\cdot x_0^3,
$$
\noindent
with the six parameters $\beta_1,\dots,\beta_6$. Following (\ref{transfmatrix}), the transfer matrix  $\cL_{\mathcal{H}}$ indexed by the states of $\mathcal{A}_1^3$ is the following:
$$
\cL_{\mathcal{H}}= \pare{
\begin{array}{cccccccc}
1 & e^{\beta_1} & e^{\beta_2} & e^{\beta_1+\beta_2+\beta_4}& e^{\beta_3}& e^{\beta_1+\beta_3+\beta_5}& e^{\beta_2+\beta_3+\beta_6}& e^{\beta_1+\beta_2+\beta_3+\beta_4+\beta_5+\beta_6}\\
\vdots & \vdots & \vdots &\vdots & \vdots & \vdots & \vdots &\vdots\\

1 & e^{\beta_1} & e^{\beta_2} & e^{\beta_1+\beta_2+\beta_4}& e^{\beta_3}& e^{\beta_1+\beta_3+\beta_5}& e^{\beta_2+\beta_3+\beta_6}& e^{\beta_1+\beta_2+\beta_3+\beta_4+\beta_5+\beta_6}
\end{array}
}
$$
\noindent
This matrix is primitive, and the unique maximum eigenvalue is 
$$\rho(\beta) =1 + e^{\beta_1} + e^{\beta_2} + e^{\beta_1+\beta_2+\beta_4}+ e^{\beta_3}+ e^{\beta_1+\beta_3+\beta_5}+ e^{\beta_2+\beta_3+\beta_6}+ e^{\beta_1+\beta_2+\beta_3+\beta_4+\beta_5+\beta_6}.$$
\noindent
The right eigenvector associated to this eigenvalue has all the components equal to 1. We obtain the topological pressure $\p{\mathcal{H}}=\log \rho(\beta)$. In order to find the MEMC parameters we solve this set of equations:

\noindent
\begin{equation} \label{eqr2}
 \frac{\partial \p{\mathcal{H}}}{\partial \beta_1}=A_T(f_{k}) =c_k.
\end{equation}
\noindent

From equation \eqref{eqr2} provided some constraints on the average value of the features we can solve the maximum entropy problem. Take for example (see table 2):

\begin{table}[h!]
\centering
\caption{ }
\label{my-label}
    \begin{tabular}{| l | l | l | l | l |}
    \hline
    $A_T(f_{k})$ & $c_{k}$ & $\beta_{k}$ & $\delta \beta_{k}$ & $\tilde{c}_{k}$ \\ \hline
   $A_T(x^1)  $ & 0.3 & -1.0436 & 0 & 0.30350016\\ \hline
     $A_T(x^2)  $  & 0.2 & -1.6727  & 0 & 0.20127414\\ \hline
    $A_T(x^3)  $  & 0.1  & -2.8163 & 0  & 0.10450018 \\ \hline
 $A_T(x^1 x^2) $  & 0.08  & 0.4590 & 0 & 0.08187418 \\ \hline
  $A_T(x^1 x^3) $  & 0.05  & 0.8604 & 0.1 &  0.05475019 \\ \hline
   $A_T(x^2 x^3) $  & 0.04  & 1.0325 & 0 & 0.04207419 \\ \hline
    \end{tabular}
\end{table}

\begin{figure}[h!]
  \centering
    \includegraphics[width=0.9\textwidth]{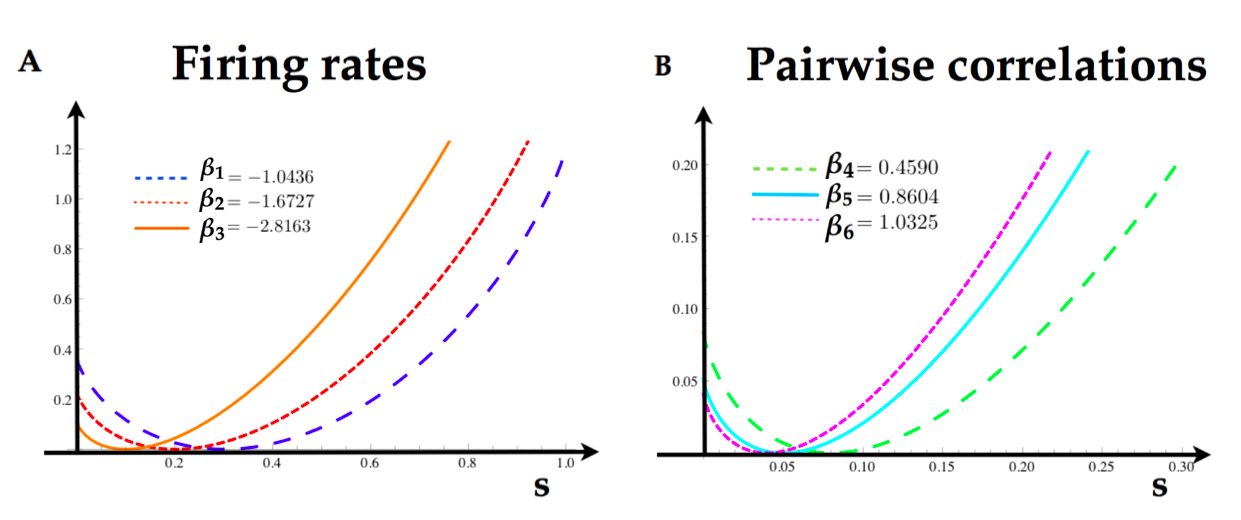}
    \caption{ \textbf{A)} Rate functions for the firing rates of each neuron of the Ising model. The minimum of the rate functions coincide with the expected value of the firing rates in the table 2. \textbf{B)} Rate functions for the pairwise interactions computed from the parameters in the table 2.}
    \label{fig:rfis}
\end{figure}

From equation \eqref{pmar} one can find that the Markov transition matrix.
%\begin{equation} \label{pisi}
%P=\frac{1}{\rho(\beta)}\cL_{\mathcal{H}}.
%\end{equation}
In order to compute the rate function of each feature in this model, we take the logarithm of the maximum eigenvalue of the tilted matrix, and obtain the tilted cumulant generating function $\lambda_f(k)$. In figure \ref{fig:rfis}) we illustrate the rate functions for each feature in the model.

\section{Conclusion} 

In the past few years, new experimental techniques combined with clever ideas from statistical mechanics have made possible to infer maximum entropy models of spike trains directly from experimental recordings. However, a very important issue which is to quantify the accuracy of the estimation obtained from a finite empirical sample is usually ignored in this field. This is probably because the maximum entropy approach has a dual nature; one side is a convex optimization problem which provides a unique solution independent of the sampling size, and on the other hand is a Bayesian inference procedure, from which is more natural to ask this question. As we have discussed in the introduction this characterization is relevant in the field of computational neuroscience as, in practice, experimental recordings are performed during a finite amount of time which causes fluctuations over the estimated quantities.

A fundamental goal of spike train analysis over networks of sensory neurons involves building accurate statistical models that predict the response of the network to a stimulus of interest. In particular, the aim of statistical inference of spiking neurons using the MEP, is that the fitted parameters shed light on some aspects of the neuronal code, therefore it is extremely important to quantify the accuracy of the statistical procedure. Additionally, one may be interested in measuring some properties of the inferred statistical model characterizing the spiking neuronal network. For example about convergence rate of a sample or to quantify the probability of rare events of features like firing rates, pairwise correlations, triplets or spatiotemporal events, mainly because these features are likely to play an important role in neuronal information processing. It is possible that rare and unlikely events have been generated by internal states of the neuronal tissue and not driven by the external stimulus. The events that are unlikely to occur deserve a better understanding as may carry important information about the network internal structure and may play a role in organizing a coherent dynamic to convey sensory information to the cerebral cortex.

%This fact calls for a better understanding of the limits of maximum entropy inference procedure and for devising ways to quantify bounds on accuracy due to finite size samplings. 
The present contribution addressed this issue using tools from large deviations theory in the context of the MEMC. In particular, we showed that the transfer matrix technique used to build the MEMC is well adapted to compute large deviation rate functions using the G\"artner-Ellis theorem. We also provide tools to investigate how sharply determined are the parameters of a MEMC with respect to the amount of empirical data using the concept of $\epsilon$ distinguishability. Additionally, we present a non-trivial relation between the distance in the parameter space and the distance in the manifold of maximum entropy probability measures using a LDP.

%Among the many ideas rooted in statistical physics that have been suggested to characterize the collective activity in the brain, perhaps the most intriguing is the idea that biological systems poise themselves at a critical point in the parameter space, which suggests that there may be some deeper theoretical principle behind this collective behavior \cite{mora-bialek:11}.
%\textcolor{blue}{esta ultima parte y los dos siguientes parrafos me suenan mas para introduccion que para conclusion...}

%How good is the fitted probability distribution for describing the statistics of the data? \cite{schneidman-berry-etal:06,vasquez-palacios-etal:12}. How likely is to generate a rare sample of a spike train from this distribution? Can we discriminate on the basis of a finite sample if two maximum entropy processes are different? \cite{balasubramanian:97}. \\

We have illustrated our method using simple examples. However, these examples might give a false impression that large deviations rate functions can always be calculated explicitly. In fact, exact and explicit expressions can be found only in small simple cases, fortunately there exist numerical methods to evaluate rate functions \cite{touchette:12}. 

Here, we have focused our attention on large deviations properties on maximum entropy models arising from spike train statistics, however, these results can be used in other fields of applications of maximum entropy models.

%%%%%%%%%%%%%%%%%%%%%%%%%%%%%%%%%%%%%%%%%%

%%%%%%%%%%%%%%%%%%%%%%%%%%%%%%%%%%%%%%%%%%
\vspace{6pt} 

%%%%%%%%%%%%%%%%%%%%%%%%%%%%%%%%%%%%%%%%%%
%% optional
%\supplementary{The following are available online at www.mdpi.com/link, Figure S1: title}

%%%%%%%%%%%%%%%%%%%%%%%%%%%%%%%%%%%%%%%%%%
\textbf{Acknowledgements} We thank Ruben Herzog and Adrian Palacios for providing us with the retinal data and for helping in figure 6. RC was supported by an ERC advanced grant "Bridges", CONICYT-PAI Inserci\'{o}n 79160120 and Proyecto REDES ETAPA INICIAL, Convocatoria 2017 REDI170457. CM was at the early stage of this project, supported by the CONICYT-FONDECYT Postdoctoral Grant No. 3140572. FR acknowledges the support of the European Union's H2020 research and innovation programme under the Marie Sk\l{}odowska-Curie grant agreement No. 702981.\\

The following abbreviations are used in this manuscript:\\

\noindent 
\begin{tabular}{@{}ll}
MEP & Maximum entropy principle\\
MEMC & Maximum entropy Markov chain\\
SCGF & Scaled cumulant generating function\\
CLT & Central limit theorem\\
LLN & Law of large numbers\\
LDP & Large deviation principle\\
IEP & Information entropy production\\
KSE & Kolmogorov-Sinai entropy\\
NESS & Non-equilibrium steady states\\
\end{tabular}

\vspace{5pt}
\noindent
\textbf{Symbol list}

\noindent 
\begin{tabular}{@{}ll}
$x_{n}^{k}$ & Spiking state of neuron $k$ at time $n$.\\
$\boldsymbol{x}_{n}$ & Spike pattern at time $n$\\
$\boldsymbol{x}_{t_1,t_2}$ & Spike block from time $t_1$ to $t_2$.\\
$A_T(f)$ & Empirical Average value of the feature $f$ considering $T$ spike patterns.\\
$\mathcal{A}_R^N$ & Set of spike blocks of $N$ neurons and length $R$.\\ 
$\mathcal{S}\bra{p}$ & Entropy of the probability measure $p$.\\
$\mathcal{H}$ & Energy function.\\
$\p{\mathcal{H}}$  & Free energy or topological pressure.\\
$\lambda_f(k)$  & Scaled cumulant generating function of $f$.\\
$I_f(s)$  & Rate function of $f$.\\
\end{tabular}

%%%%%%%%%%%%%%%%%%%%%%%%%%%%%%%%%%%%%%%%%%
%% optional
%\appendixtitles{yes} %Leave argument "no" if all appendix headings stay EMPTY (then no dot is printed after "Appendix A"). If the appendix sections contain a heading then change the argument to "yes".
%\appendixsections{multiple} %Leave argument "multiple" if there are multiple sections. Then a counter is printed ("Appendix A"). If there is only one appendix section then change the argument to "one" and no counter is printed ("Appendix").
%\appendix
%\section{}
%\subsection{}
%The appendix is an optional section that can contain details and data supplemental to the main text. For example, explanations of experimental details that would disrupt the flow of the main text, but nonetheless remain crucial to understanding and reproducing the research shown; figures of replicates for experiments of which representative data is shown in the main text can be added here if brief, or as Supplementary data. Mathematical proofs of results not central to the paper can be added as an appendix.

%\section{}
%All appendix sections must be cited in the main text. In the appendixes, Figures, Tables, etc. should be labeled starting with `A', e.g., Figure A1, Figure A2, etc. 

\appendix

\section{Discrete-time Markov chains and spike train statistics}\label{Appen-Markov-Spike}

Consider the random process $\{X_{n}:n\geq0\}$ taking values on $\mathcal{A}_{R}^{N}$. One can assume that the spiking activity of the neuronal network can be modeled by some discrete-time Markov process whose transition probabilities are obtained by means of the maximum entropy method described in section~\ref{sec:3}. In this setting, $\mathcal{A}_{R}^{N}$ is the state space of the Markov chain, and thus, if $X_{n} = x_{n,n+R-1}$ we say that the process is in the state $x_{n,n+R-1}$ at time $n$. The transition probabilities are given as follows, 
\begin{equation}\label{Markov1}
\mathbb{P}\big[X_{n}=x_{(n)} \mid X_{n-1}=x_{(n-1)}, \hdots, X_{0}=x_{(0)}\big] = \mathbb{P}\big[X_{n}=x_{(n)} \mid X_{n-1}=x_{(n-1)}\big],
\end{equation}
where we used the short hand notation $x_{(n)}:= x_{n,n+R-1}$. We emphasize that in this paper the states are spike blocks of finite length $R$, $x_{n,n+R-1}$. All along this paper he have only considered homogeneous Markov chains, that is,~\eqref{Markov1} is independent of $n$.

Since transitions are considered between blocks of the form $x_{n-R,n-1} \rightarrow x_{n-R+1,n}$, therefore the block $x_{n-R+1,n-1}$ must be common for the transition to be possible. Consider two spike blocks $i,j \in \mathcal{A}_R^N$ of range $R\geq 2$. We say that the transition from  state $i$ to state $j$ is \textit{allowed} if $i$ and $j$ have the common sub-block $x_{1,R-1}= \tilde{x}_{0,R-2}$, where $ \tilde{x}_{0,R-2}$ are the first $R-1$ spike patterns of $j$.

Now, we define the transition matrix $P_{R}:\mathcal{A}_R^N\times\mathcal{A}_R^N\to \setR$, whose entries are given by the transition probabilities, as follows,

\begin{equation}\label{transmatrix}
(P_{R})_{ij}:= 
\left\{
\begin{array}{lll}
[j \mid i] > 0 
\quad &\mbox{if }   i \to j  \mbox{ is allowed } \\
0, \quad &\mbox{otherwise}.
\end{array}
\right.
\end{equation}
\noindent
Note that $P$ has $ 2^{NR} \times 2^{NR}$ entries, but it is a sparse matrix since each line has, at most, $2^N$ non-zero entries. A stochastic matrix $P$ is defined from transition probabilities~\eqref{transmatrix} satisfying:
\begin{equation*}%\label{sumu}
\mathbb{P}[j \mid i] \geq 0; \quad \quad \sum_{j \in \mathcal{A}_R^N} \mathbb{P}[j \mid i] =1,
\end{equation*}
for all states $i,j  \in \mathcal{A}_R^N$. Moreover, by construction, for any pair of states, there exists a path of maximum length $R$ in the graph of transition probabilities going from one to the other, which means that the Markov chain is primitive.

\section{Cumulant generating function}\label{Appen-cum-gen}

In general in order to obtain the scale cumulant generating function (as considered in section~\ref{sec:ld} ) one has to deal with the moment of order $r$ of a real-valued random variable $f$, which is given by,
\[
m_r=\mathbb{E}(f^r),
\]
for $r \in \mathbb{N}$. Provided that it has a Taylor expansion about the origin, the moment generating function (or Fourier-Laplace transform) 
\[
M(k) = \mathbb{E}(e^{k f}) 
= \mathbb{E}(1 + k f +\cdots + k^r f^r/r!+\cdots) 
	= \sum_{r=0}^\infty m_r k^r/r!
\]
The cumulants $\kappa_r$ are the coefficients in the Taylor expansion of the cumulant generating function, defined as the logarithm of the moment generating function, that is,
$$\log M(k) = \sum_{r} \kappa_r k^r/r!$$
The relationship between the first moments and cumulants, can be obtained by extracting coefficients from the expansion, as follows:

$$
\begin{array}{lcl}
\kappa_1 &=& m_1 \\
\kappa_2 &=& m_2 - m_1^2\\
\kappa_3 &=& m_3 - 3m_2m_1 + 2m_1^3\\
\kappa_4 &=& m_4 - 4m_3m_1 - 3m_2^2 + 12m_2m_1^2 -6m_1^4,
\end{array}
$$
\noindent
and so on. In particular, $\kappa_1$ is the mean of $f$ , $\kappa_2$ is the variance, $\kappa_3$ the skewness and $\kappa_4$ the kurtosis. 

\section{Linear response}\label{lin-rep}

Within the framework we have build we can quantify how a small perturbation of the maximum entropy parameters (associated to given features) affects the average values of other features of the MEMC. It is important to quantify this perturbation because the maximum entropy parameters are obtained with finite accuracy due to finite sample effects. Fixing $\boldsymbol{\beta}$, we can obtain the average value of a given feature $f_k$ with respect to the MEMC without need to sample, using the Gibbs-Jaynes principle for the KSE~\cite{georgii:03}, which asserts that for a translation invariant probability measure $p$, the entropy rate $S_{KS}(p)$ is maximal under the constraints $\mathbb{E}_p\{f_k\} = c_k$, for all $k \in \{1,\dots,K\}$ if and only if $p$ is a Gibbs measure associated to the energy $\mathcal{H}_{\beta}=\sum \beta_k f_k$, where $\mathbb{E}_p\{f_k\}=\frac{\partial \mathcal{P}[\mathcal{H}_{\beta}]
}{\partial \beta_k}=c_k$.\\

%\begin{equation} \label{eqrr}
%\mu_{\boldsymbol{\beta}}[f_k]=\frac{\partial \p{\mathcal{H}_{\boldsymbol{\beta}}}}{\partial \beta_k}=c_k
%\end{equation}

Now, let us consider a perturbed version of the energy denoted by $\mathcal{H}_{\boldsymbol{\beta} + \delta \boldsymbol{\beta}}$. Using a Taylor expansion, we compute the average value of an arbitrary feature here denoted by $f_k$ with respect to the MEMC associated to the perturbed energy in terms of the unperturbed one, that is,

\begin{eqnarray}\label{lr1}
\mathbb{E}_{p_{\boldsymbol{\beta} + \delta \boldsymbol{\beta} }}\{f_k\} &=& \frac{\partial \p{\mathcal{H}_{\boldsymbol{\beta} + \delta \boldsymbol{\beta} }}}{\beta_k} \\
  &=&  \frac{\partial \p{\mathcal{H}_{\boldsymbol{\beta}}}}{\beta_k} +  \sum_j \frac{\partial^2 \p{\mathcal{H}_{\boldsymbol{\beta}}}}{\partial \beta_k \beta_j }\delta \beta_j + O(\delta \beta_j)^2 \label{lr2}\\
    &=&  \mathbb{E}_{p_{\beta}}\{f_k\}+  \sum_j \frac{\partial^2 \p{\mathcal{H}_{\boldsymbol{\beta}}}}{\partial \beta_k \beta_j }\delta \beta_j + O(\delta \beta_j)^2=\tilde{c}_k. \label{lr3}
\end{eqnarray} 
\noindent
From (\ref{lr1}) to (\ref{lr2}) there is a Taylor expansion of $\p{\mathcal{H}_{\boldsymbol{\beta} + \delta \boldsymbol{\beta}}}$ about $\mathcal{H}_{\boldsymbol{\beta}}$. From (\ref{lr2}) to (\ref{lr3}) we use the Gibbs-Jaynes principle for the KSE. We see from (\ref{lr3}) that a small perturbation of a parameter $\beta_j$ influence the average value of all other features in the energy function (as $f_k$ is arbitrary) and the magnitude of the perturbation is controlled by the second derivatives of the topological pressure of the unperturbed energy $\p{\mathcal{H}_{\boldsymbol{\beta}}}$. 

\section{Time Correlations from Topological Pressure}\label{tc-tp}

%For a first-order stationary Markov chain, since each $X_n, n\geq 1$ depends on its predecessor, this induces a non-zero time-correlation between
%$X_n$ and $X_{n+r}$, even when the distance $r$ is greater than 1. This correlation is termed the autocorrelation of order $r$. 

For a pair of finite range features $f_k,f_j$ of a stationary Markov chain, the covariance of order $r$ is independent of time, just depend on the lag $r$ and is defined as:

$$C_{f_k,f_j}(r):=\mathbb{E}_{p}\left\lbrace f_k (x_n) f_j(x_{n+r})\right\rbrace - \mathbb{E}_{p}\left\lbrace f_k (x_n)\right\rbrace \mathbb{E}_{p}\left\lbrace  f_j(x_{n}) \right\rbrace , $$

\noindent
where $\mathbb{E}_{p}$ stands for the expected value with respect to the Markov measure $p$.  

%In particular, the autocovariance of order $r$ of the $f_k$ reads as:

%$$C_{f_k}(r):=\mathbb{E}_{\mu}\left[ f_k (x_n) f_k(x_{n+r})\right] - \mathbb{E}^2_{\mu}\left[ f_k (x_n)\right]  $$

%For an ergodic Markov we expect the autocorrelation of any finite range observable to decrease as the lag is increased.
\noindent
For MEMC with potentials of range $R>1$ there is a positive time correlation correlation between pairs of features $f (x_n)$ and $ g(x_{n+r})$, that we denote $\sigma_{f,g}^2$, indeed one can show that (Green-Kubo formula):

\beq\label{gcfo}
\sigma_{f_k,f_j}^2=C_{f_k,f_j}(0)+\sum_{r=1}^{\infty} C_{f_k,f_j}(r)+\sum_{r=1}^{\infty} C_{f_j,f_k}(r).
\eeq

%\noindent
%In particular, the autocorrelation is:

%$$\sigma_{f_k}^2=C_{f_k}(0)+2\sum_{r=1}^{\infty} C_{f_k}(r)$$

\noindent
The pairwise time correlations between features can be obtained from the topological pressure:

\beq\label{eqap2}
\sigma_{f_k,f_j}^2=\frac{\partial^2  \mathcal{P}[\mathcal{H}_{\beta}]}{\partial \beta_k \, \partial \beta_j}=\frac{\partial \moy{f_j}}{\partial \beta_{k}}.
\eeq

\noindent
For a MEMC taking values on $\mathcal{A}_R^N$ characterized by $p(\pi, P)$ : 
$$
\frac{\partial^2  \mathcal{P}[\mathcal{H}_{\beta}]}{\partial \beta_k \, \partial \beta_j} = \mathbb{E}_p\{f_k f_j\}-\mathbb{E}_p\{f_k\}\mathbb{E}_p\{f_j\}+ \sum_{r=1}^{\infty} \sum_{y,w \in \mathcal{A}_R^N } f_k(y)f_j(w)\pi_{y}P^{r}_{yw} + \sum_{r=1}^{\infty} \sum_{y,w  \in \mathcal{A}_R^N} f_j(y)f_k(w)\pi_{y}P^{r}_{yw}$$

\noindent
For $v,w \in \mathcal{A}_R^N$. For MEMC fitted through range one energy functions $\{f (x_n);n\geq 0\}$ is an i.i.d. process and the variance of $f$ is simply $C_f(0)$. These terms are the linear response coefficients. For MEMC associated to energy functions formed by $K$ features, the matrix $L$ can be conveniently arranged in a $K \times K$ symmetric matrix (known as the Onsager reciprocity relations \cite{gaspard:14}).

\beq\label{eqap3}
\sigma_{f_k,f_j}^2=\frac{\partial^2  \mathcal{P}[\mathcal{H}_{\beta}]}{\partial \beta_k \, \partial \beta_j}=\frac{\partial \mathbb{E}_p\{f_j\}}{\partial \beta_{k}}.
\eeq

%\subsubsection{Gallavotti-Cohen fluctuation theorem}%\label{varpro}
\section{Gallavotti-Cohen fluctuation theorem}\label{gc-ft}

The Gallavotti-Cohen fluctuation theorem refers to a universal property of the IEP i.e. is independent of the parameters of the MEMC. It is as a statement about properties of the SCGF and rate function of the IEP \cite{jiang:04}, this is, 

\begin{equation} \label{gcsr}
\lambda_W(k)=\lambda_W(-k-1), \quad I_W(s) =I_W(-s) - s.
\end{equation}

This symmetry can be seen as a generalization of Kubo formula (\ref{gcfo}) and Onsager reciprocity relations (\ref{eqap3}) to situations far from equilibrium. It is a relationship that holds for a general class fs stochastic processes including Markov chains \cite{maes:99}. 

These properties have an impact on the large deviations of the time-averaged entropy production rate of the sample trajectory $\boldsymbol{x}_{0,t-1}$ of the Markov chain $p(\pi,P)$ denoted $\frac{W_t}{t}$. In our framework, the following relationship always holds, 
\[
\frac{p \left\lbrace\frac{W_t}{t} \approx s \right\rbrace}{p \left\lbrace\frac{W_t}{t} \approx -s\right\rbrace } \asymp e^{ts}.
\]

\noindent
This means that the positive fluctuations of $\frac{W_t}{t}$ are exponentially more probable than negative fluctuations of equal magnitude.

\bibliographystyle{abbrv}
\bibliography{bibliothesis31}

\end{document}